\documentclass{article}[12pt]
\usepackage[large,bf,FIGTOPCAP]{subfigure}
\usepackage{lineno}
\usepackage{setspace}
\usepackage{jneurosci}
\usepackage{graphicx}
\usepackage{color}
\usepackage{epsfig}
\usepackage{amsmath, amsthm, amsfonts}
\usepackage{amssymb}
\usepackage{latexsym}
\usepackage[draft]{hyperref}

\begin{document}


\vspace{.5cm}

\begin{center}
{\LARGE Constructing precisely computing networks with biophysical spiking neurons}\\
\vspace{.5cm}
{\large Michael A. Schwemmer$^{1,\dagger}$, Adrienne L. Fairhall$^{3,5*}$, Sophie Den{\'e}ve$^{2*}$ and \\ Eric T. Shea-Brown$^{4*}$}\\
\vspace{.3cm}
{$^1$ Mathematical Biosciences Institute, The Ohio State University, Columbus, Ohio 43210}\\
{$^2$ Laboratoire de Neurosciences Cognitives, INSERM, CNRS, {\'E}cole Normale Sup{\'e}rieure,\\ Paris, France 75005}\\
{$^3$ Department of Physiology and Biophysics and $^4$ Department of Applied Mathematics,}\\
{$^5$ WRF UW Institute for Neuroengineering and Data Sciences Senior Fellow,} \\
{University of Washington, Seattle, Washington 98195}\\
{$^*$ equal contribution.}
\end{center}

\noindent $^{\dagger}$ Corresponding Author: Mathematical Biosciences Institute, The Ohio State University, Jennings Hall 3rd Floor, 1735 Neil Ave., Columbus, OH 43210; schwemmer.2@mbi.osu.edu \\

\noindent {\bf Acknowledgements:} MS is supported in part by the Mathematical Biosciences Institute and the National Science Foundation under grant DMS 0931642.  AF acknowledges support from NSF grant 0928251. ESB acknowledges support from a Simons Fellowship in Mathematics, and from NSF grant 1122106, and is grateful for the hospitality of the Allen Institute for Brain Science, where he is currently a visiting scientist.  SD acknowledges support from 
ANR-10-LABX-0087 IEC, ANR-10-IDEX-0001-02 PSL, ERC grant FP7-PREDISPIKE, and a James McDonnell Foundation Award - Human Cognition. This work was supported in part by an allocation of computing time from the Ohio Supercomputer Center. 

\newpage


%

\section*{Abstract}

While spike timing has been shown to carry detailed stimulus information at the sensory periphery, its possible role in network computation is less clear.  
Most models of computation by neural networks are based on population firing rates. In equivalent spiking implementations, firing is assumed to be random such that averaging across populations of neurons recovers the rate-based approach. 
Recently, however, Den{\'e}ve and colleagues have suggested that the spiking behavior of neurons may be fundamental to how neuronal networks compute, with precise spike timing determined by each neuron's contribution to producing the desired output. By postulating that each neuron fires in order to reduce the error in the network's output, it was demonstrated that linear computations can be carried out by networks of integrate-and-fire neurons that communicate through instantaneous synapses. This left open, however, the possibility that realistic networks, with conductance-based neurons with subthreshold nonlinearity and the slower timescales of biophysical synapses, may not fit into this framework. Here, we show how the spike-based approach can be extended to biophysically plausible networks. We then show that our network reproduces a number of key features of cortical networks including irregular and Poisson-like spike times and a tight balance between excitation and inhibition.  Lastly, we discuss how the behavior of our model scales with network size, or with the number of neurons ``recorded" from a larger computing network. These results significantly increase the biological plausibility of the spike-based approach to network computation.


\section*{Significance Statement}

We derive a network of neurons with standard spike-generating currents and synapses with realistic timescales that computes based upon the principle that the precise timing of each spike is important for the computation.  We then show that our network reproduces a number of key features of cortical networks including irregular, Poisson-like spike times and a tight balance between excitation and inhibition. These results significantly increase the biological plausibility of the spike-based approach to network computation, and uncover how several components of biological networks may work together to efficiently carry out computation.


\section*{Introduction}

Neural networks transform their inputs through a variety of computations from the integration of stimulus information for decision-making \cite{gold2007} to the persistent activity observed in working memory tasks~\cite{jonides2008}. How such transformations occur in biological networks has not yet been understood. Such operations have been proposed to be carried out by the averaged firing rates of neurons in a network (a ``rate model") \cite{goldmanetal2003,machensetal2005,seung1996,wang2002,wongandwang2006}. 
 For example, persistent activity may be realized in a rate model by including recurrent connections that balance the intrinsic leak of the system \cite{goldman2009,seung1996}.  However, most real neural circuits consist of spiking neurons.  Spiking network implementations of rate model operations can be constructed by assuming that the computation is distributed among a large population of functionally similar neurons, so that the averaged firing rate matches that of the desired rate model \cite{eckhoff2011,renartetal2004,wongandwang2006}. 

Rate-based approaches have been used to model a variety of behaviors including persistent activity in the oculomotor integrator \cite{goldmanetal2003,seung1996,seungetal2000}, decision-making \cite{eckhoff2011,boga03,wongandwang2006,UMcCl,CainSB12}, and working memory \cite{renartetal2004,BrodyRK03}. While rate models capture features of both psychophysical and electrophysiological data, such approaches have a few potential limitations. First, any rate-based approach disregards the timing of individual spikes, and hence any capacity to compute that precise timing may confer. Second, the performance of rate models is typically quite sensitive to the choice of connection weights between neural populations \cite{seungetal2000}.  If the recurrent connections are either too strong or too weak, the activity of the network can either quickly increase to saturation or decrease to a baseline level. Further, spiking network implementations of rate-based networks typically { (though not always, e.g. \cite{limandgoldman2013})} require strong added noise to match the irregular firing observed in cortical networks \cite{machensetal2005,wang2002,wongandwang2006}. This injected noise often dominates the feed-forward or intrinsic currents generated in the network, diminishing the accuracy with which inputs can be integrated or maintained over time.

Recently, Boerlin et al. (2013) have proposed a distinct alternative by assuming that a computation is carried out directly by the spiking times of individual neurons. Based upon the premise that the membrane potentials of neurons in the network track a prediction error between a desired output and the network estimate, and that neurons spike only if that error exceeds a certain value, Boerlin et al. (2013) derived a spiking neural network that can perform any linear computation.  In this {\em predictive coding} approach, the computation error is mapped to the voltage of integrate-and-fire (IF) neurons, while a bound on this error is mapped to the neuron's threshold. This leads to a recurrent network of IF neurons with a mixture of instantaneous and exponential synapses that is able to reproduce many features of cortical circuits while performing a variety of linear computations including pure and leaky integration, differentiation, and transforming inputs into damped oscillations. Furthermore, as the computation is efficiently distributed among the participating neurons, the network is robust to perturbations such as lesions and synaptic failure.

Nevertheless, two components of this work potentially limit its implementation in biological networks:  neurons communicate instantaneously, while true synaptic dynamics occur with a finite timescale; and the threshold of IF neurons is set arbitrarily, rather than being established by intrinsic nonlinear spike-generating kinetics.

Here, starting from the same spike-based framework \cite{boerlin2013}, we derive a computing network of neurons with standard spike-generating currents~\cite{hodgkinhuxley1952} and synapses with realistic timescales.  Like in many cortical networks, the spike times of the model network are irregular and there is a tight balance between excitation and inhibition \cite{okunandlampl2008,renartetal2010,shadlen1998}.  Moreover, the precise timing of spikes is important for accurate decoding:  the network actively produces correlations in the spike times of different neurons which act to reduce the decoding variance.  Taken together, the results uncover how several components of biological networks may work together to efficiently carry out computation.


\section*{Materials and Methods}

\subsection*{Optimal Spike-based Computation with Finite Time-scale Synapses} \label{optimalnet}

Here, we follow Boerlin et al. (2013) to construct a spiking network that implements the computation of a $J$-dimensional linear dynamical system. We define the target system as

\begin{equation}
\dot{\mathbf{x}} = A\mathbf{x}+\mathbf{\bar{c}}(t) \label{xeqn},
\end{equation}

\noindent where $\mathbf{x}(t)$ is a $J$-dimensional vector of functions of time, $\mathbf{\bar{c}}(t)$  is a $J$-dimensional vector of stimulus inputs, and $A$ is a $J\times J$ matrix (with units of $s^{-1}$) that determines the linear computation.  For example, if $A$ is the zero matrix, then the computation would be pure integration with $\mathbf{x}(t)$ being the integral of the stimulus inputs  $\mathbf{\bar{c}}(t)$. The dynamic variables $\mathbf{x}$ are unitless while time has units of seconds.  We want to build a network of $N$ neurons such that an estimate of the dynamic variable $\mathbf{\hat{x}}\approx\mathbf{x}$ can be read out from the network's spike trains $\rho_k(t)=\sum_j\delta(t-t_j^k)$, where $k$ indexes the $N$ neurons.  We assume that the dynamics of the network estimate $\mathbf{\hat{x}}$ are given by:

\begin{equation*}
\mathbf{\dot{\hat{x}}}=-a_d\mathbf{{\hat{x}}}+ \Gamma \boldsymbol{\rho}\ast h_r(t),
\end{equation*}

\noindent where $h_r(t)=(a_r-a_d)/\alpha^*H(t)e^{-a_r t}$, $H(t)$ is the Heaviside function, $a_r$, $a_d$, and $\alpha^*$ are constants that are defined below, and $\Gamma$ is a $J\times N$ dimensional decoding matrix. In the original LIF network \cite{boerlin2013}, $h_r(t)=\delta(t)$.  The solution of the above equation, assuming $\mathbf{\hat{x}}(0)=\mathbf{0}$, is given by the convolution of the network's spike trains with a  double-exponential function

\begin{equation}
\mathbf{\hat{x}}(t)=\Gamma \boldsymbol{\rho}\ast \alpha(t)=\int_0^t\Gamma\boldsymbol{\rho}(s)\alpha(t-s)ds,
\end{equation}

\noindent where

\begin{equation}
\alpha(t) = H(t)\frac{1} {\alpha^*} (e^{-a_d t} - e^{-a_r t}) \label{alphafun},
\end{equation}

\noindent and $\alpha^*$ is a constant so that the maximum of the double-exponential function is $1$, $a_r$ ($a_d$) is the rate of rise (decay) of the double-exponential function.  Note that the normalization term ${(a_r-a_d)/\alpha^*}$ in the definition of ${h_r(t)}$ comes from the fact that we wanted $\alpha(t)$ to have the form given above. In what follows, we will show that this alteration to the decoder dynamics will result in a neuronal network with finite time-scale synapses.

We now derive network dynamics such that neurons spike in order to reduce the error between the signal $\mathbf{x}(t)$ and the estimate $\mathbf{\hat{x}}(t)$.  Defining the error function $E(t)$ as

\begin{equation}
E(t)=\int_0^{t}\left(\sum_{j=1}^J(x_j(u)-\hat{x}_j(u))^2\right)du, \label{ErrorFun}
\end{equation}

\noindent our goal is to derive conditions under which cell $k$ spikes only if the error is reduced by doing so: $E(t|\textrm{cell k spikes}) < E(t|\textrm{cell k doesn't spike})$.   When cell $k$ spikes at time $t$, this changes $\hat{x}_j(u) \to \hat{x}_j(u)+\Gamma_{jk}\alpha(u-t)$.  Thus, we need to find conditions such that

\begin{equation*}
\int_0^{t}\left(\sum_{j=1}^J(x_j(u)-\hat{x}_j(u)-\Gamma_{jk}\alpha(u-t))^2\right)du < \int_0^{t}\left(\sum_{j=1}^J(x_j(u)-\hat{x}_j(u))^2\right)du.
\end{equation*}

\noindent Up to this point, our derivation is nearly identical to that of Boerlin et al. (2013), except for the use of the double-exponential function synapse.  However, we must now alter the above condition in order to account for the fact that the double-exponential function synapse has a finite rise time.  More specifically, since $\alpha(t)$ is equal to zero at the time of the spike, the terms on either side of the above inequality are equal (since $\alpha(u-t)=0$ for $u\leq t$). In contrast, Boerlin et al. used exponential synapses which have an infinitely fast rise time and thus yield a non-zero contribution at the time of a spike. Thus, in order to account for the effects of the spike at time $t$ on the error, we need to extend the integration a short time $t^*$ into the future:

\begin{equation*}
\int_0^{t+t^*}\left(\sum_{j=1}^J(x_j(u)-\hat{x}_j(u)-\Gamma_{jk}\alpha(u-t))^2\right)du < \int_0^{t+t^*}\left(\sum_{j=1}^J(x_j(u)-\hat{x}_j(u))^2\right)du.
\end{equation*}

\noindent After some algebra, and using the fact that $\alpha(u-t)=0$ for $u\leq t$, this leads to

\begin{equation*}
\int_t^{t+t^*}\left(\sum_{j=1}^J2\Gamma_{jk}\alpha(u-t)(x_j(u)-\hat{x}_j(u))\right)du > \int_t^{t+t^*}\left(\sum_{j=1}^J\Gamma_{jk}^2\alpha^2(u-t)\right)du.
\end{equation*}

\noindent Since $t^*$ is assumed to be small, we approximate the above integral using the trapezoidal rule

\begin{equation*}
\frac{1} {2} \left(\sum_{j=1}^J2\Gamma_{jk}(\alpha(0)(x_j(t)-\hat{x}_j(t))+\alpha(t^*)(x_j(t+t^*)-\hat{x}_j(t+t^*)))\right)t^* > \frac{1} {2}\left(\sum_{j=1}^J\Gamma_{jk}^2(\alpha^2(0)+\alpha^2(t^*))\right)t^*.
\end{equation*}

\noindent Other integral approximations lead to similar results.  Using the fact that $t^*$ is small, we can Taylor expand $x_j(t+t^*)$ and $\hat{x}_j(t+t^*)$ to first order

\begin{equation*}
\frac{1} {2} \left(\sum_{j=1}^J2\Gamma_{jk}\alpha(t^*)\left[x_j(t)-\hat{x}_j(t)+(x'_j(t)-\hat{x}'_j(t))t^*\right]\right)t^* > \frac{1} {2}\left(\sum_{j=1}^J\Gamma_{jk}^2\alpha^2(t^*)\right)t^*,
\end{equation*}

\noindent where we used the fact that $\alpha(0)=0$. Dividing both sides of the above equation by $\alpha(t^*)t^*$ we arrive at

\begin{equation*}
\sum_{j=1}^J\Gamma_{jk}(x_j(t)-\hat{x}_j(t))+\sum_{j=1}^J\Gamma_{jk}(x_j'(t)-\hat{x}_j'(t))t^*  > \sum_{j=1}^J\frac{\Gamma_{jk}^2} {2}\alpha(t^*).
\end{equation*}

\noindent Lastly, we drop terms of order $t^*$ and define

\begin{equation*}
\tilde{V}_k = \sum_{j=1}^J\Gamma_{jk}(x_j(t)-\hat{x}_j(t)),
\end{equation*}

\noindent with the condition that neuron $k$ spikes when it reaches threshold $T_k=\sum_{j=1}^J\frac{\Gamma_{jk}^2} {2}\alpha(t^*)$.

The network dynamics are given by differentiating the above equation

\begin{eqnarray*}
\dot{\tilde{V}}_k &=&\Gamma_k^T(\mathbf{\dot{x}}-\mathbf{\dot{\hat{x}}}) \\
&=&\Gamma_{k}^T\left(A\mathbf{{x}}(t)+\mathbf{\bar{c}}(t)-\dot{\hat{\mathbf{x}}}(t)\right).
\end{eqnarray*}

\noindent To close the problem using only information available to the network, we replace the desired signal with the spike-based {\em estimate} of the signal, $\mathbf{x}(t)\approx \mathbf{\hat{x}}(t)$:

\begin{eqnarray*}
\dot{\tilde{V}}_k &\approx&\Gamma_{k}^T(A\mathbf{\hat{x}}(t)+\mathbf{\bar{c}}(t)-\dot{\hat{\mathbf{x}}}(t))\\
&=& \Gamma_{k}^T(A+a_d\mathbb{I}_{J\times J})\Gamma\boldsymbol{\rho}\ast\alpha(t)+\Gamma_{k}^T\mathbf{\bar{c}}(t) - \Gamma_{k}^T\Gamma\boldsymbol{\rho}\ast h_r(t)
\end{eqnarray*}

\noindent The above form highlights the fact that there are now two different kinds of synapses in our network: double-exponential ``slow'' synapses and exponential ``fast'' synapses. The reason why these two types of synapses arise is because both $\mathbf{\hat{x}}(t)$ and its temporal derivative appear in the equation for the voltage dynamics. If we had chosen to decode the spike trains using an exponential kernel as in \cite{boerlin2013}, we would end up with exponential slow synapses and $\delta$-function fast synapses.

In previous approaches, the neurons' voltage ``reset" following spikes arose from autaptic (i.e., from a neuron to itself) input currents via the delta-function synapses just discussed.  Such fast synapses do not occur in our derivation.  To obtain an analogous reset condition, we would need to include an additional, explicit reset current in our voltage equation.  This would result in:

\begin{equation*}
\dot{\tilde{V}}_k = \Gamma_{k}^T(A+a_d\mathbb{I}_{J\times J})\Gamma\boldsymbol{\rho}\ast\alpha(t)+\Gamma_{k}^T\mathbf{\bar{c}}(t)- \Gamma_{k}^T\Gamma\boldsymbol{\rho}\ast h_r(t)-2 T_k \rho_k(t),
\end{equation*}

\noindent where the term $-2 T_k \rho_k(t)$ resets neuron $k$ to $-T_k$ once it reaches threshold $T_k$.  We illustrate this particular reset rule because it matches that of \cite{boerlin2013}.  However, in the next section we will remove this reset term and replace it with more biologically realistic ionic currents.

Next, we rescale the voltage to be in terms of $mV$ (recall that time is in units of seconds). To do so, we introduce the scaling $\tilde{V_k}=\frac{T_k} {g}V_k$ (where $g$ has units of $mV$) which leads to

\begin{equation*}
\dot{V_k}= -2 g \rho_k(t) + \frac{g} {T_k}\left[\Gamma_{k}^T(A+a_d\mathbb{I}_{J\times J})\Gamma\boldsymbol{\rho}\ast\alpha(t)+\Gamma_{k}^T\mathbf{\bar{c}}(t)- \Gamma_{k}^T\Gamma\boldsymbol{\rho}\ast h_r(t)\right],
\end{equation*}

\noindent where the threshold voltage is $g$ and the reset voltage is $-g$.  The parameter $g$ also modifies the gain of the synaptic input. However, it is also linked to the value of the voltage threshold and reset potential. Finally, to frame the network equations in terms of current, we multiply both sides by the membrane capacitance $C_m$ (in units of $mF/cm^2$)

\begin{equation*}
C_m\dot{V_k}= -2C_m g \rho_k(t) + \frac{C_mg} {T_k}\left[\Gamma_{k}^T(A+a_d\mathbb{I}_{J\times J})\Gamma\boldsymbol{\rho}\ast\alpha(t)+\Gamma_{k}^T\mathbf{\bar{c}}(t)-\Gamma_{k}^T\Gamma\boldsymbol{\rho}\ast h_r(t)\right].
\end{equation*}


\subsection*{Addition of Biophysical Currents}

We began by deriving a network of neurons that do not contain any intrinsic biophysical currents and solely integrate their synaptic input prior to spiking. To incorporate the nonlinear dynamics of spike-generating ion channels, we now replace the reset currents $-2 C_m g \rho_k(t)$ with generic Hodgkin-Huxley-type (HH-type) ionic currents $I_{ion}(V_k,\vec{w}_k)$ (see section \nameref{params} for a specific example)

\begin{equation*}
C_m\dot{V_k}= I_{ion}(V_k,\vec{w}_k) + \frac{C_mg} {T_k}\left[\Gamma_{k}^T(A+a_d\mathbb{I}_{J\times J})\Gamma\boldsymbol{\rho}\ast\alpha(t)+\Gamma_{k}^T\mathbf{\bar{c}}(t)- \Gamma_{k}^T\Gamma\boldsymbol{\rho}\ast h_r(t)\right],
\end{equation*}

\noindent where the $\vec{w}_k$ in $I_{ion}(V_k,\vec{w}_k)$ represent the gating variables for standard HH currents.  For example, $\vec{w}_k=[m_k,h_k,n_k]$ for the HH-type model we consider here (see section \nameref{params}).  For simplicity, we assume that every neuron in the network has the same type of spike-generating currents $I_{ion}(V_k,\vec{w}_k)$.  Note that if we wanted to use a leaky-integrate-and-fire neuron, we would set $I_{ion}(V_k,\vec{w}_k)=-g_L(V_k-E_L)-2gC_m \rho_k(t)$, where $g_L$ is the conductance of the leak channel (in $mS/cm^2$), $E_L$ is the leak channel reversal potential, and we used the same reset current we previously described. As stated above, for standard HH-type model currents there is no longer a need for a reset current as the spiking process is carried out by the intrinsic currents.

Next, we add a white noise current to our voltage equations.  This is meant to roughly model a combination of background synaptic input, randomness in vesicle release, and stochastic fluctuations in ion channel states (channel noise), but also contributes to computation in our networks by helping to prevent synchrony (see Results section \nameref{paramsens}).  The result is:  

\begin{equation*}
C_m\dot{V_k}= I_{ion}(V_k,\vec{w}_k) + \frac{C_mg} {T_k}\left[\Gamma_{k}^T(A+a_d\mathbb{I}_{J\times J})\Gamma\boldsymbol{\rho}\ast\alpha(t)+\Gamma_{k}^T\mathbf{\bar{c}}(t)-\Gamma_{k}^T\Gamma\boldsymbol{\rho}\ast h_r(t)\right]+\sigma_V\xi_k(t),
\end{equation*}

\noindent where $\xi(t)$ is white noise ($\langle\xi(t)\rangle =0$ and $\langle \xi(t)\xi(t')\rangle = \delta(t-t')$) { and $\sigma_V$ has units of $\mu A/cm^2 \cdot \sqrt{s}$}. Lastly, to emphasize the fact that the input to the system $\mathbf{\bar{c}}(t)$ has the physical interpretation of current, we introduce the scaling $\mathbf{\bar{c}}(t) = \mathbf{c}(t) /(C_m c_0)$ where $\mathbf{c}(t)$ has units of $\mu A/cm^2$ and $c_0$ has units of $mV$ and scales the stimulus input into neurons in our network.  Thus, we rewrite the above equation as

\begin{equation*}
C_m\dot{V_k}= I_{ion}(V_k,\vec{w}_k) + \frac{C_mg} {T_k}\left[\Gamma_{k}^T(A+a_d\mathbb{I}_{J\times J})\Gamma\boldsymbol{\rho}\ast\alpha(t)-\Gamma_{k}^T\Gamma\boldsymbol{\rho}\ast h_r(t)\right]+\frac{\Gamma_{k}^T} {T_k}\frac{g} {c_0}\mathbf{c}(t)+\sigma_V\xi_k(t).
\end{equation*}

\noindent Switching to vector notation, the population dynamics are given by

\begin{equation}
C_m \dot{\mathbf{V}}= \mathbf{I_{ion}}(\mathbf{V},\vec{\mathbf{w}}) + C_m g \tilde{T}^{-1}\left[\Gamma^T(A+a_d\mathbb{I}_{J\times J})\Gamma\boldsymbol{\rho}\ast\alpha(t)- \Gamma^T\Gamma\boldsymbol{\rho}\ast h_r(t)\right]+\tilde{T}^{-1} \Gamma^T \frac{g} {c_0}\mathbf{c}(t)+\sigma_V{\boldsymbol \xi(t)},
\end{equation}

\noindent where $\tilde{T}$ is an $N\times N$ diagonal matrix with $T_k$ on the diagonal.

In the integrate-and-fire network, spiking occurs due to an explicit threshold crossing and reset condition.  With the addition of ionic currents, action potentials are now intrinsically generated, but it is still necessary to identify a voltage to identify spike times.  We treat this detection threshold as a separate parameter.  In the simulation presented, we chose to use $V_{th}=-48$ $mV$ which is sufficiently high on the upswing of the action potential to allow reliable spike detection. However, different choices for $V_{th}$ can lead to different behaviors for the network.  In particular, our simulations show that in order to use a larger value for $V_{th}$, one must also increase the voltage noise in order to prevent the network from synchronizing.


\subsection*{Compensating for Spike-generating Currents}\label{derivekernels}

In the previous section, we incorporated spike-generating currents into the voltage dynamics of each cell in our network.  The point of this is to add biological realism, but the immediate consequence is that the voltages no longer evolve to precisely track error signals for the intended computation.  This degrades the accuracy with which the network can perform.  However, in this section we show that it is possible to effectively ``compensate'' the network for the effects of the spike-generating currents.  

To begin, we note that, assuming no noise, a network optimized for the underlying computation maintains the relationship

\begin{equation}
\mathbf{V}=g\tilde{T}^{-1}\Gamma^T(\mathbf{x}-\mathbf{\hat{x}}), \label{Vrel}
\end{equation}

\noindent i.e., the voltage of each cell represents a projection of the error signal. However, the addition of the spike-generating currents disrupts the relationship (\ref{Vrel}).  Thus, we seek to derive alterations to both the network and decoder dynamics in order to make (\ref{Vrel}) valid.  That is, we take the dynamics of $\mathbf{V}$ and $\mathbf{\hat{x}}$ to be given by

\begin{eqnarray}
&&\dot{\mathbf{V}} = \frac{\mathbf{I_{ion}}(\mathbf{V},\vec{\mathbf{w}})} {C_m} + \mathbf{I}(t) \label{newV}\\
&&\mathbf{\dot{\hat{x}}}=-a_d\mathbf{\hat{x}}+\Gamma \boldsymbol{\rho}\ast h_r(t)+\mathbf{G}(\mathbf{V}) \label{newxh}
\end{eqnarray}

\noindent where $\mathbf{I}(t)$ and $\mathbf{G}(\mathbf{V})$ are functions to be determined in order to restore the relationship between voltage and error, Eq. (\ref{Vrel}). Taking the derivative of equation (\ref{Vrel}) and using (\ref{newxh}), we find

\begin{eqnarray*}
\dot{\mathbf{V}} &=&g \tilde{T}^{-1}\Gamma^T(\mathbf{\dot{{x}}}-\mathbf{\dot{\hat{x}}})\\
&=&g \tilde{T}^{-1}\Gamma^T\left(A\mathbf{x}+\frac{\mathbf{c}(t)} {C_m c_0}+a_d\mathbf{\hat{x}}-\Gamma \boldsymbol{\rho}\ast h_r(t)-\mathbf{G}(\mathbf{V})\right)\\
&\approx&g \tilde{T}^{-1}\left[ \Gamma^T(A+a_d \mathbb{I}_{J\times J})\mathbf{\hat{x}}+\Gamma^T\frac{\mathbf{c}(t)} {C_mc_0}-\Gamma^T\Gamma\boldsymbol{\rho}\ast h_r(t)\right]-g \tilde{T}^{-1}\Gamma^T\mathbf{G}(\mathbf{V}),
\end{eqnarray*}

\noindent where above we again used the fact that $\mathbf{x}\approx \mathbf{\hat{x}}$. Equating this definition of the derivative of $\mathbf{V}$ to (\ref{newV}), we find

\begin{eqnarray*}
&&\mathbf{I}(t)=g \tilde{T}^{-1}\left[ \Gamma^T(A+a_d \mathbb{I}_{J\times J})\mathbf{\hat{x}}+\Gamma^T\frac{\mathbf{c}(t)} {C_mc_0}-\Gamma^T\Gamma\boldsymbol{\rho}\ast h_r(t)\right]\\
&&\mathbf{G}(\mathbf{V})=-\Phi\frac{\mathbf{I_{ion}}(\mathbf{V},\vec{\mathbf{w}})} {C_m g},
\end{eqnarray*}

\noindent where $\Phi=\left(\Gamma^T\right)^{\dagger}\tilde{T}$ and $\left(\Gamma^T\right)^{\dagger}=(\Gamma \Gamma^T)^{-1} \Gamma$ is the Moore-Penrose pseudoinverse of the rectangular matrix $\Gamma^T$. Thus, the new dynamics would be

\begin{eqnarray*}
&&\dot{\mathbf{V}} = \frac{\mathbf{I_{ion}}(\mathbf{V},\vec{\mathbf{w}})} {C_m} + g \tilde{T}^{-1}\left[ \Gamma^T(A+a_d \mathbb{I}_{J\times J})\mathbf{\hat{x}}-\Gamma^T\Gamma\boldsymbol{\rho}\ast h_r(t)\right] +\tilde{T}^{-1}\Gamma^T\frac{g} {c_0}\frac{\mathbf{c}(t)} {C_m}\\
&&\mathbf{\dot{\hat{x}}}=-a_d\mathbf{\hat{x}}+\Gamma\boldsymbol{\rho}\ast h_r(t)-\Phi\frac{\mathbf{I_{ion}}(\mathbf{V},\vec{\mathbf{w}})} {C_m g},
\end{eqnarray*}

\noindent which implies that $\mathbf{V}$ and $\mathbf{\hat{x}}$ are coupled, as the solution of $\mathbf{\hat{x}}$ is (ignoring initial conditions):

\begin{equation}
\mathbf{\hat{x}}(t)=\int_0^t\Gamma\boldsymbol{\rho}(s)\alpha(t-s)ds - \int_0^t\Phi\frac{\mathbf{I_{ion}}(\mathbf{V}(s;\mathbf{\hat{x}}(s)),\vec{\mathbf{w}}(s))} {C_m g}e^{-a_d(t-s)}ds.\label{xhat1}
\end{equation}

\noindent  This coupling implies that the decoder  $\mathbf{\hat{x}}$ requires instantaneous knowledge of the voltages of each cell.  Clearly, a more realistic -- and simpler -- implementation would be if the decoder had access only to the spike times of the cells.   We next show how this can be achieved.  We begin with the assumption that the primary cause of the disruption of (\ref{Vrel}) occurs only during an action potential. We then find an approximation of the intrinsic current $I_{ion}(V,\vec{w})/C_m$ that follows a spike.  That is, we seek a kernel $\eta(t)$ such that

\begin{equation}
\frac{I_{ion}(V_k(t),\vec{w})} {C_m}\approx \int_{t_j^k}^{t_j^k+t_s} \rho_k(s)\eta(t-s)ds,
\end{equation}

\noindent where $t_j^k$ is the time of the $j^{th}$ spike of cell $k$ and $t_s$ is the width of the kernel $\eta(t)$.  More details on obtaining the kernel $\eta(t)$ are provided in the next section.  Plugging the above approximation into the last term in Eq. (\ref{xhat1}), we obtain

\begin{eqnarray}
\mathbf{\hat{x}}(t)&\approx& \int_0^t\Gamma\boldsymbol{\rho}(s)\alpha(t-s)ds - \frac{1} {g} \Phi\boldsymbol{\rho}\ast\left[\int_0^t\eta(s)e^{-a_d(t-s)}ds\right] \nonumber\\
&=& \Gamma\boldsymbol{\rho}\ast\alpha(t) - \frac{1} {g} \Phi\boldsymbol{\rho}\ast\tilde{\eta}(t), \label{xhatapp}
\end{eqnarray}

\noindent where $\tilde{\eta}(t)= \int_0^t\eta(s)e^{-a_d(t-s)}ds$.  Note that $\eta(t)$ has units of $mV/s$ while $\tilde{\eta}(t)$ has units of $mV$.  We can then rewrite the network dynamics as

\begin{eqnarray*}
&&\dot{\mathbf{V}} = \frac{\mathbf{I_{ion}}(\mathbf{V},\vec{\mathbf{w}})} {C_m} + g \tilde{T}^{-1}\left[ \Gamma^T(A+a_d \mathbb{I}_{J\times J})\Gamma\boldsymbol{\rho}\ast\alpha(t)-\Gamma^T\Gamma\boldsymbol{\rho}\ast h_r(t)\right]\\
&& \hspace{2cm}-\tilde{T}^{-1}\Gamma^T(A+a_d \mathbb{I}_{J\times J})\Phi\boldsymbol{\rho}\ast\tilde{\eta}(t)+\tilde{T}^{-1}\Gamma^T\frac{g} {c_0}\frac{\mathbf{c}(t)} {C_m}+\frac{\sigma_V} {C_m} {\boldsymbol \xi}(t)\\
&&\mathbf{\dot{\hat{x}}}=-a_d\mathbf{\hat{x}}+\Gamma\boldsymbol{\rho}\ast h_r(t)-\frac{1} {g}\Phi\boldsymbol{\rho}\ast \eta(t),
\end{eqnarray*}

\noindent where the voltage noise term has again been included. Finally, we introduce the following more compact notation:

\begin{eqnarray}
&&\dot{\mathbf{V}} = \frac{\mathbf{I_{ion}}(\mathbf{V},\vec{\mathbf{w}})} {C_m} + g\Omega_s\boldsymbol{\rho}\ast\alpha(t)-g\Omega_f\boldsymbol{\rho}\ast h_r(t)-\Omega_c\boldsymbol{\rho}\ast\tilde{\eta}(t)+\frac{g} {c_0}\frac{1} {C_m}D\mathbf{c}(t) +\frac{\sigma_V} {C_m} {\boldsymbol \xi}(t) \\
&& \mathbf{\hat{x}}(t) = W\ast\boldsymbol{\rho}(t),
\end{eqnarray}

\noindent where	

\begin{eqnarray*}
&& \Omega_s =  \tilde{T}^{-1}\Gamma^T(A+a_d \mathbb{I}_{J\times J})\Gamma  \\
&& \Omega_f =  \tilde{T}^{-1}\Gamma^T\Gamma  \\
&& \Omega_c =  \tilde{T}^{-1}\Gamma^T(A+a_d \mathbb{I}_{J\times J})\Phi \\
&& D =  \tilde{T}^{-1}\Gamma^T  \\
&& W(t) = \alpha(t) \Gamma - \frac{\tilde{\eta}(t)} {g} \Phi.
\end{eqnarray*}

\noindent We reiterate that the compensation affects both the network dynamics and the read-out.  Note also that the parameter $g$ scales the strength of the slow and fast synaptic input.


\subsection*{Obtaining the Compensation Kernels}\label{getkernels}

The compensation kernel $\eta(t)$ was obtained by stimulating a single model neuron with a Gaussian noise current (specifically, an Ornstein-Uhlenbeck process~\cite{uhlenbeck1930}), and keeping track of the times $t_j$ that the voltage crossed a threshold from below. This threshold was the same as that used for detecting spikes in the network simulations. For each spike, we then obtain an action potential waveform $V_{AP}^j(t)$ for $t_j\leq t < t_j+t_s$, where $t_s$ sets the width of the $\eta(t)$ kernel.  We then sum these traces to obtain the average waveform of the action potential $V_{AP}(t)$.  That is, if $K$ spikes were recorded, then

\begin{equation*}
V_{AP}(t)=\frac{1} {K} \sum_{j=1}^{K} V_{AP}^j(t-t_j), \,\,\,\,\, 0\leq t< t_s.
\end{equation*}

\noindent Thus, an approximation to the change in voltage during the spike is given by

\begin{equation*}
\frac{I_{ion}(V,\vec{w})} {C_m} \approx \frac{d} {dt} V_{AP}(t), \,\,\,\,\, 0\leq t< t_s.
\end{equation*}

\noindent The kernel $\eta(t)$ is then defined as

\begin{equation}
\eta(t) =
\begin{cases}
\frac{d} {dt} V_{AP}(t) &  0\leq t < t_s \\
0 & \textrm{ otherwise}.
\end{cases}
\end{equation}

\noindent See Figure \ref{CompEx} for an illustration of this procedure.  For our simulations, we set $t_s=4$ $ms$.  Using a larger value of $t_s$ did not significantly affect the results, but too small a value does, as the voltage trace during the entire time course of the action potential will not be accounted for.


\subsection*{Decoding Variance and Approximations} \label{decvarsec}

In this section, we assume that the network tracks a one-dimensional signal; that is, $J=1$.  The decoder is given by

\begin{equation*}
\hat{x}= \Gamma\boldsymbol{\rho}^{\alpha} - \frac{1} {g} \Phi\boldsymbol{\rho}^{\tilde{\eta}},
\end{equation*}

\noindent where $\boldsymbol{\rho}^Y = \boldsymbol{\rho}\ast Y(t)$, $Y\in\{\alpha, \tilde{\eta}\}$.  The variance of the decoder is then given by

\begin{equation}
\textrm{var}(\hat{x})\equiv \nu_{\hat{x}}=\Gamma C^{\alpha} \Gamma^T + \frac{1} {g^2}\Phi C^{\tilde{\eta}}\Phi^T -\frac{2} {g} \Gamma C^{\alpha \tilde{\eta}} \Phi^T, \label{decvar}
\end{equation}

\noindent where $C^{\alpha}_{ij}=\textrm{cov}(\rho^{\alpha}_i,\rho^{\alpha}_j)$, $C^{\tilde{\eta}}_{ij}=\textrm{cov}(\rho^{\tilde{\eta}}_i,\rho^{\tilde{\eta}}_j)$, and $C^{\alpha \tilde{\eta}}_{ij}=\textrm{cov}(\rho^{\alpha}_i,\rho^{\tilde{\eta}}_j)$. Similarly, the variance of a decoder that assumes that all neurons are independent is given by
\begin{equation}
\nu^{\textrm{ind}}_{\hat{x}}=\Gamma D^{\alpha} \Gamma^T + \frac{1} {g^2}\Phi D^{\tilde{\eta}}\Phi^T -\frac{2} {g} \Gamma D^{\alpha \tilde{\eta}} \Phi^T, \label{decvarind}
\end{equation}

\noindent where $D^X$ shares the same diagonal elements with $C_X$ but is zero on the off-diagonals and $X=\{\alpha, \tilde{\eta}, \alpha \tilde{\eta}\}$.  

In the main text we quantify the relative decoding variance of the independent vs. ``full" (i.e., correlated) network via the fraction $\nu^{\textrm{ind}}_{\hat{x}}/\nu_{\hat{x}}$.  Values of this fraction greater than one indicate that the network produces correlated spike times that reduce decoding variance vs. the ``shuffled," independent case; we refer to it as the ``reduction in decoding variance."  To compute this quantity, we performed $800$ two-second runs of the network, with a new noise realization on each trial, calculated the covariance matrices for each trial, averaged the covariance matrices across all trials and used the averaged matrices in Eqs. (\ref{decvar}) and (\ref{decvarind}).

For the homogeneous network considered below, we can obtain a simple estimate for the reduction in decoding variance. Suppose that $\Gamma_k=a$ for $k=1,..,N/2$ (stimulus activated population; see main text) and $\Gamma_k=-a$ for $k=N/2+1,...,N$ (stimulus depressed population; see main text) for some constant $a$.  Then $\Phi_k=b$ for $k=1,..,N/2$ and $\Phi_k=-b$ for $k=N/2+1,...,N$ for some constant $b$ related to $a$.  Assume that the variance of each neuron is very close to the average variance over the population, i.e., that the diagonals of each of the above covariance matrices are constant.  Dividing each of the above covariance matrices by this average variance yields a matrix with ones on the diagonal and the various pairwise correlation coefficients on the off-diagonals. Assuming that the pairwise correlation coefficients are close to their average values, the above matrices have a very simple form:

$$
C^{X}=\sigma_X
\begin{bmatrix}
1 & a_{X} & a_X & ... &  c_{X} & c_X & ... & c_{X} \\
a_X & 1 & a_X  & ... &  c_X  & c_X &  ... & c_X \\
.   &   &   .  &     &      &    &   & . \\
.   &   &     &  .   &       &    &   & . \\
c_X & c_X &  ... & c_X & 1  & d_X & ... & d_X \\
.   &   &      &     &      &  .  &   & . \\
.   &   &     &      &       &    & .  &  \\
c_X & c_X &  ... & c_X & d_X  & ... & d_X & 1 \\
\end{bmatrix},
$$

\noindent where $a_{X}$ ($d_{X}$) is the mean correlation coefficient for the stimulus activated (stimulus depressed) population computed using kernel $X$, and $c_{X}$ is the mean correlation coefficient between the two different populations using kernel $X$. With this approximation, the elements of the above variance calculations take a simple form

\begin{eqnarray*}
&& \Gamma C^{\alpha} \Gamma^T = \sigma_{\alpha} a^2\left[N + \frac{N^2-2N} {4} (a_{\alpha} + d_{\alpha}) -\frac{N^2} {2} c_{\alpha}\right]\\
&& \Phi C^{\tilde{\eta}} \Phi^T = \sigma_{\tilde{\eta}} b^2\left[N + \frac{N^2-2N} {4} (a_{\tilde{\eta}} + d_{\tilde{\eta}}) -\frac{N^2} {2} c_{\tilde{\eta}}\right]\\
&& \Gamma C^{\alpha\tilde{\eta}} \Phi^T = \sigma_{\alpha\tilde{\eta}} a^2\left[N + \frac{N^2-2N} {4} (a_{\alpha\tilde{\eta}} + d_{\alpha\tilde{\eta}}) -\frac{N^2} {2} c_{\alpha\tilde{\eta}}\right],
\end{eqnarray*}

\noindent and

\begin{eqnarray*}
&& \Gamma D^{\alpha} \Gamma^T = \sigma_{\alpha} a^2N\\
&& \Phi D^{\tilde{\eta}} \Phi^T = \sigma_{\tilde{\eta}} b^2N\\
&& \Gamma D^{\alpha\tilde{\eta}} \Phi^T = \sigma_{\alpha\tilde{\eta}} a b N.
\end{eqnarray*}

\noindent Thus, an approximation to the reduction in decoding variance obtained by recording from only a subset of the full network is given by using the above formulae in $\nu^{\textrm{ind}}_{\hat{x}}/\nu_{\hat{x}}$, since the correlation coefficients do not vary with $N$.  However, if we assume that the dominant contribution to the variance calculation is given by those terms involving the $C_{\alpha}$ matrix (which is what we find numerically, cf. Fig. \ref{recsubsetfig} (a)), then an even simpler formula can be obtained

\begin{equation*}
\frac{\nu^{\textrm{ind}}_{\hat{x}}} {\nu_{\hat{x}}} \approx \frac{1} {1-\frac{a_{\alpha}+d_{\alpha}} {2}+N\frac{a_{\alpha}+d_{\alpha}-2c_{\alpha}} {4}},
\end{equation*}


\subsection*{Computing Correlation Coefficients}\label{computingCC}

The reported correlation coefficients between cells $i$ and $j$ are computed by convolving spike trains with a double-exponential function, so that $\boldsymbol{\rho}^{\alpha} = \boldsymbol{\rho}\ast \alpha(t)$:

\begin{equation*}
CC_{ij}=\frac{\sum_{n=1}^T(\rho^{\alpha}_i(t_n)-\bar{\rho}^{\alpha}_i)(\rho^{\alpha}_j(t_n)-\bar{\rho}^{\alpha}_j)} {\sqrt{\sum_{n=1}^T(\rho^{\alpha}_i(t_n)-\bar{\rho}^{\alpha}_i)^2}\sqrt{\sum_{n=1}^T(\rho^{\alpha}_j(t_n)-\bar{\rho}^{\alpha}_j)^2}},
\end{equation*}

\noindent where $T$ is the total number of time points taken for a given simulation, $t_n$ is the $n$-th time point, and $\bar{\rho}^{\alpha}=T^{-1}\sum_{n=1}^T\rho^{\alpha}(t_n)$ is the sample mean. To remove the covariance in firing rates of the cells, the correlation coefficients were corrected by subtracting off the correlation coefficient obtained from shift-predictor data (shifted by one trial). Since our networks consist of two populations of neurons, i.e., those with a positive value for $\Gamma$ and those with a negative value for $\Gamma$, the correlation coefficients reported in the histograms are the population-averaged correlation coefficients for each trial simulation of the network.  To generate the histograms, we ran $800$ two-second simulations of the network with the same box function input. The only thing that varied between the simulations was the realization of the white background noise.


\subsection*{Computing Fano Factors}\label{Fano}

The Fano factors for each neuron were computed by binning the spike times into $20$ $ms$ windows and computing the mean $\mu_w$ and variance $\sigma_w^2$ of the spike count in a particular window over $800$ repeated trials of the box function input stimulus.  The Fano factor in a particular window is then given by $\sigma^2_w /\mu_w$.  For each neuron, the time averaged Fano factor was computed by taking the mean over all windows.  We then averaged these values over all neurons in a given population and report them in Figure \ref{ISIFig}.


\subsection*{Error Metrics}\label{errormetrics}

Two measures of error quantify the network performance. The first is the relative error between the signal and the estimate,

\begin{equation}
\frac{||x-\hat{x}||_2} {||x||_2},
\end{equation}

\noindent where $||f||_2 = \sqrt{\int_0^T[f(s)]^2ds}$ and $T$ is the simulation time.  Relative error is useful for comparing errors across signals that vary in magnitude.  The second error measure is the integrated squared error,

\begin{equation}
\int_0^T[x(s)-\hat{x}(s)]^2 ds.
\end{equation}


\subsection*{Voltage Cross-correlograms and Power Spectra}\label{Vcorr}

To analyze the subthreshold voltages of cells in our network, we first truncated the membrane potentials at $-60$ $mV$ to remove the spikes and subtracted out the temporal mean, i.e., $\bar{V}_m(t) = V_m(t)-\frac{1} {N} \sum_{t=1}^N V_m (t)$ where $N$ is the total number of data points. Voltage power spectra for individual neurons were then computed using Matlab's \emph{fft} function. Cross-correlations between two cells $\bar{V}_{m1}$  and $\bar{V}_{m2}$ were also calculated using Matlab's \emph{xcorr} function:

\begin{equation}
R_{12}(\tau) = \frac{\sum_{t=1}^{N-\tau}\bar{V}_{m1}(t+\tau)\bar{V}_{m2}(t)} {\sqrt{\sum_{t=1}^{N}\bar{V}_{m1}^2(t)\sum_{t=1}^{N}\bar{V}_{m2}^2(t)}} ,\,\,\,\, \tau\geq 0; \,\,\,\, R_{12}(\tau)=R_{21}(-\tau),\,\,\, \tau<0,
\end{equation}
where $\tau$ is the time lag \cite{lampl1999,yu2010}. We then subtracted off the cross-correlation for shift-predictor data (shifted by one trial).  Both the power spectra and cross-correlograms were then averaged over $1,000$ eight-hundred millisecond simulations of the homogeneous integrator network with the box function input.


\subsection*{Computing the Spike-triggered Error Signal}\label{STEcomp}

The spike-triggered error of Figure \ref{ISIFig} was computed from $800$ two-second simulations of the network with a box function input (see below).  For each simulation, we computed

\begin{equation*}
\mathbf{e}(t)=\Gamma^T(x(t)-\hat{x}(t)),
\end{equation*}

\noindent where $\mathbf{e}\in \mathbb{R}^{N}$ is the nondimensional error each neuron is supposed to be representing in its voltage traces.  The error $e_k(t)$ was aligned to the spike times for cell $k$ and these traces averaged over all neurons in the network.  The shuffled spike-triggered error, computed by aligning $e_k(t)$ to the spike times of cell $k$ on a different trial, was then subtracted. This removed the slow bias present in the original spike-triggered error.  Lastly, the shuffle-corrected spike-triggered errors were averaged over all trials.


\subsection*{Measuring Population Synchrony} \label{popsync}

The level of synchrony in the simulated network was evaluated using a measure introduced by Golomb \cite{golomb2007}. With $f_k(t)$ as the instantaneous firing rate of neuron $k$, synchrony is given by

\begin{equation}
\chi^2(N)=\frac{\left\langle \left[\frac{1} {N}\sum_{k=1}^N f_k(t)\right]^2\right\rangle_t - \left[ \left\langle\frac{1} {N}\sum_{k=1}^N f_k(t)\right\rangle_t\right]^2} {\frac{1} {N} \sum_{k=1}^N\left\langle \left[f_k(t)\right]^2\right\rangle_t - \left[ \left\langle f_k(t)\right\rangle_t\right]^2},
\end{equation}

\noindent where $\langle ... \rangle_t$ denotes time-averaging over the length of the simulation.  To estimate instantaneous firing rates, the spike trains were convolved with a gaussian kernel with standard deviation $10$ $ms$.


\subsection*{Scaling when Varying the Simulated Network Size} \label{VaryNscaling}

When varying the simulated network size as in Figure \ref{VaryNFig}, we scaled the connection strengths of the network so that the total input to any cell in the network remains constant as the network size is increased.  In particular, for the homogeneous integrator ($A=0$) network tracking a one-dimensional dynamical system where $\Gamma_k = a$ for $k=1,2,...,N/2$ and $\Gamma_k = -a$ for $k=N/2+1,...,N$, we employed the scaling:

\begin{eqnarray*}
&& a=\frac{40} {N} \\
&& g=c_0 \frac{400} {N}.
\end{eqnarray*}

\noindent Thus, both the connection weights and the synaptic gain parameter $g$ scale with $1/N$.  The factors of $40$ and $400$ above were chosen so that at $N=400$, $\Gamma_k = \pm 0.1$ and $c_0=g$, which matches our earlier simulations of our network when we fixed $N$ at $400$. With this scaling, the connection strengths all scale the same way with $N$ and the input $c(t)$ remains constant.  To see this, recall that when $A=0$ our network equations are given by

\begin{eqnarray*}
&&\dot{V_k} = \frac{I_{ion}(V_k,\vec{w}_k)} {C_m} + \frac{g} {T_k}\left[a_d\Gamma_k^T\Gamma \boldsymbol{\rho}\ast \alpha(t) - \Gamma_k^T\Gamma \boldsymbol{\rho}\ast h_r(t)+\Gamma_k^T \frac{c(t)} {C_mc_o}\right]-a_d\frac{1} {T_k}\Gamma_k^T\Phi \boldsymbol{\rho}\ast \tilde{\eta}(t) + \frac{\sigma_V} {C_m} \xi_k(t),
\end{eqnarray*}

\noindent where $T_k=\frac{\Gamma_k^2} {2}$, $\Phi=\left(\Gamma^T\right)^{\dagger}\tilde{T}$, and $\tilde{T}$ is a diagonal matrix with $T_k$ on the diagonal. Thus, we need to determine the scaling of the following:

\begin{eqnarray*}
&& \textrm{Slow Connections: } a_d\frac{g} {T_k} \Gamma_k^T\Gamma  \\
&& \textrm{Fast Connections: } \frac{g} {T_k} \Gamma_k^T\Gamma \\
&& \textrm{Compensating Connections: } a_d\frac{1} {T_k}\Gamma_k^T\Phi \\
&& \textrm{Feedforward Weights: } g \frac{\Gamma_k^T} {T_k}  \\
&& \textrm{Noise (external input): } \sigma_V.
\end{eqnarray*}

\noindent First, we explore the term $\Phi$ which involves the pseudoinverse of the $\Gamma^T$ matrix. In the case of the homogeneous integrator network, the pseudoinverse is simply given by $\left(\Gamma^T\right)^{\dagger}_k=\pm \frac{1} {Na}$, because $\left(\Gamma^T\right)^{\dagger} \Gamma^T = 1$. Thus, if we let $ a \sim 1/N$ as listed above, $\left(\Gamma^T\right)^{\dagger}_k \sim 1$. $\Phi_k=\left[\left(\Gamma^T\right)^{\dagger}\tilde{T}\right]_k$ then scales like $1/N^2$.  Using this fact, and recalling that $g\sim 1/N$, we can now compute the scalings for all the connections in the network:

\begin{eqnarray*}
&& \textrm{Slow Connections: } a_d\frac{g} {T_k} \Gamma_k^T\Gamma \sim (1) (N^2/N) (1/N) (1/N) = \frac{1} {N} \\
&& \textrm{Fast Connections: } \frac{g} {T_k} \Gamma_k^T\Gamma \sim (1/N) (N^2/N) (1/N) = \frac{1} {N} \\
&& \textrm{Compensating Connections: } a_d\frac{1} {T_k}\Gamma_k^T\Phi \sim (1) (N^2) (1/N) (1/N^2) = \frac{1} {N} \\
&& \textrm{Feedforward Weights: } g \frac{\Gamma_k^T} {T_k} \sim (1/N) (N^2/N) = 1 \\
&& \textrm{Noise (external input): } \sigma_V \sim 1,
\end{eqnarray*}

\noindent where we used the fact that since $a_d$ and $\sigma_V$ are constants, they scale like $1$.  Thus, the connection weights scale like $1/N$.  However, since each cell in the network receives input from all $N$ other cells, this scaling means that the total input each cell receives remains constant as the network size is varied.


\subsection*{Models and Parameters} \label{params}

We use a neuron model due to Traub et al. \cite{traubmiles1995,hoppensteadtpeskin2001}:

\begin{eqnarray*}
& & C_m \frac{d v} {d t} = - g_{Na}m^3 h (v(t)-E_{Na}) - g_K n^4(v(t)-E_K)-g_L(v(t)-E_L) \\
& & \frac {dm} {dt} = 10^3(\alpha_m(v)(1-m)-\beta_m(v)m)\\
& & \frac {dh} {dt} = 10^3(\alpha_h(v)(1-h)-\beta_h(v)h)\\
& & \frac {dn} {dt} = 10^3(\alpha_n(v)(1-n)-\beta_n(v)n)
\end{eqnarray*}

\noindent where

\begin{tabbing}
$\alpha_m(v)=1.28\frac{(v+54)/4} {1-\exp(-(v+54)/4)}$ \=\hspace{1cm} \=$\beta_m(v)=1.4\frac{(v+27)/5} {\exp(-(v+27)/5)-1}$\\
$\alpha_h(v)=0.128 \exp(-(v+50)/18)$ \> \>$\beta_h(v)=4.0\frac{1} {1+\exp(-(v+27)/5)}$\\
$\alpha_n(v)=0.16\frac{(v+52)/5} {1-\exp(-(v+52)/5)}$ \> \>$\beta_n(v)=0.5 \exp(-(v+57)/40)$
\end{tabbing}

\noindent and

\begin{tabbing}
$C_m = 10^{-3}$ $mF/cm^2$ \= \hspace{1cm} \=$g_{Na} =100$ $mS/cm^2$\= \hspace{1cm}\=$g_K = 80$ $mS/cm^2$\\
$g_L = 0.2$ $mS/cm^2$\> \>$E_{Na} = 50$ $mV$\> \> $E_{K} = -100$ $mV$\\
$E_{L} = -67$ $mV$
\end{tabbing}

\noindent The factor of $10^3$ in the gating variable equations comes from conversion of time units from milliseconds to seconds.  Other neuron models, including an exponential integrate-and-fire model, were used with similar results.

Other parameters held constant in our simulation are:

\begin{tabbing}
$a_r=200$ $Hz$ \= \hspace{1cm} \=$a_d=50$ $Hz$\\
$V_{th} = -48$ $mV$ .
\end{tabbing}

\noindent The decay rate of $a_d=50$ $Hz$ yields a decay time constant of $20$ $ms$ for the slow,  double-exponential function synapses in our network. This decay time constant is in the range of those observed in inhibitory and excitatory post synaptic currents \cite{rotaruetal2011,xiangetal1998}.  The rise rate $a_r=200$ $Hz$ sets the decay time scale for the fast, exponential synapses.  These synapses have a decay time constant of $5$ $ms$, as has been observed in inhibitory cells in rat somatosensory cortex \cite{salinetal1996}.


\subsection*{Simulations}

Simulations were written in MATLAB. The Euler-Maruyama method was used to integrate the stochastic differential equations using a time step of $0.01$ $ms$.  Simulations with time steps of $0.005$ and $0.02$ $ms$ yielded similar results.  Spikes were counted as voltage crossings of a threshold of $-48$ $mV$ from below. The initial voltages for the network were chosen randomly, while the channel variables were set to their steady-state values given the fixed initial voltage.  In particular, the initial voltages were chosen from a Gaussian distribution with a mean of $E_L$ and a standard deviation of $9 mV$. The initial state for the signal and the decoded estimate were both set to zero, i.e., $x(0)=\hat{x}(0)=0$.

Though we have provided the most general form for the network tracking any linear dynamical system, throughout the majority of the paper, we focus on the case of a homogeneous network integrating a one-dimensional signal.  That is, we set $J=1$,  $A=0$, and $\Gamma_j = a$ for $j=1,...,N/2$ and $\Gamma_j = -a$ for $j=N/2+1,...,N$, where $a$ is a constant.  The only exception to this is in the examples in Figure 1 where we set $A=-a_d$ in order to remove the slow synapses in the network dynamics.  We also set $c_0=g$ for all figures except Figure \ref{VaryNFig}.

We focus on the network integrating one of two different signals.  The first varies between two constant values (``box" input):

\begin{equation*}
c(t)=
\begin{cases}
0.08\textrm{ } \mu A/cm^2 & \textrm{ for } t_0 \leq t <  t_0+50 \\
0\textrm{ } \mu A/cm^2 & \textrm{ otherwise, }
\end{cases}
\end{equation*}

\noindent where $t_0=100$ $ms$ for Figure \ref{NetEx} and $t_0=0$ $ms$ for all subsequent figures.  The second is a frozen Ornstein-Uhlenbeck \cite{uhlenbeck1930} signal given by

\begin{equation*}
\frac{dc} {dt} = -\frac{c} {\tau} + \frac{\sigma} {\tau} \xi(t),
\end{equation*}

\noindent where $\xi(t)$ is a frozen white noise realization with zero mean and unit variance, $\tau=10$ $ms$, { and $\sigma=0.008$ $\mu A/cm^2 \cdot s^{3/2}$}.


\section*{Results}

\subsection*{Spike-based Computation with Conductance-based Neurons}

Our goal in this work is to design a network to carry out an arbitrary linear computation on an input over time --- and to do so with neurons that generate spikes via realistic ionic currents and synaptic timescales. Writing the computation as a linear dynamical system, $\dot{\mathbf{x}} = A\mathbf{x}+\rm{input}$, where $A$ is a constant matrix and $\mathbf{x}$ is the signal we desire to compute, Boerlin et al. (2013) were able to construct a recurrent spiking network to accomplish this goal.  The strategy was to arrange connections so that the voltage of each neuron would be proportional to a difference between the currently decoded network output and the ideal computation, trigger spikes when this error exceeds a threshold, and communicate these spikes (and hence the error) to other neurons in the network.  Thus, every action potential occurs at a precise time that serves to reduce the ``global" computational error across the network.  We refer to this framework as spike-based computation.

In this previous work, the authors successfully mapped the requirement of each spike reducing output error onto a network of recurrently connected linear integrate-and-fire neurons with instantaneous synaptic dynamics. However, biological networks have slower synaptic kinetics, and have ionic currents with nonlinear dynamics that determine spike generation.  Here, we will show how these two aspects of neurophysiology in fact can fit naturally with spike-based computation. 

In particular, we want to design a network of neurons such that an estimate $\mathbf{\hat{x}}(t)$ of a $J\times 1$ vector of signal variables $\mathbf{x}(t)$ can be linearly read out from the spike times of the network. As above, we assume the signal variables obey a general linear differential equation $\dot{\mathbf{x}} = A\mathbf{x}+\rm{input}$. Thus, $A$ is a $J\times J$ dimensional matrix and the input is $J$-dimensional. The entries of the matrix $A$ determine the type of computation the network is asked to perform on the $J$-dimensional inputs, which we will denote as $\mathbf{c}(t)$. For example, if $A$ is the zero matrix, then the network integrates each component of the input over time. Our network will consist of $N$ neurons with output given by the $N$ spike trains, written as $\rho_k(t)=\sum_j\delta(t-t_j^k)$ $k=1,...,N$.  

Our first goal is to incorporate synapses that have finite temporal dynamics. The synaptic dynamics enter through the definition of a {\em decoder} that provides an estimate for the variable $\mathbf{x}$.  This decoder includes a linear transformation of the network spike trains $\boldsymbol{\rho}(t)$ via a $J\times N$ linear decoding matrix $\Gamma$.  The spike trains  $\rho_k$ are first convolved with the synaptic filter $\alpha(t)$ ($\boldsymbol{\rho} \ast \alpha(t) = \int \boldsymbol{\rho}(s)\alpha(t-s)ds$), which we take to be a standard double-exponential function.  With these synaptic dynamics in this decoding, an estimate of the computed variable is given by  $\hat{\mathbf{x}}(t) = \Gamma \boldsymbol{\rho}\ast \alpha(t)$. The $\Gamma$ matrix will determine the connectivity structure of the network (see Materials and Methods section \nameref{optimalnet}).  

Given this decoder, we now follow \cite{boerlin2013} to derive the network dynamics and connectivity.  The key step is to requiring that neurons in the network only spike in order to reduce the integrated squared error between the signal and its decoded estimate.   As shown in Materials and Methods (see Eq. \ref{ErrorFun}), this has the consequence that each neuron in the network has a voltage that is equivalent to a weighted error signal, i.e., the voltage of the $k^{th}$ neuron is given by $V_k(t) \propto \Gamma_k^T(\mathbf{x}(t)-\mathbf{\hat{x}}(t))$ ($\Gamma_k^T$ is the $k$-th column of the $N\times J$ matrix $\Gamma^T$).  Each neuron then fires when its own internal copy of the error signal exceeds a set threshold value. The optimal network that carries out this spike-based computation is given by a network of ``pure integrate-and-fire" neuron models that directly integrate synaptic inputs without any leak or intrinsic membrane currents; however, a linear leakage current can be added to the voltage dynamics for each neuron with minimal disruption of the network dynamics~\cite{boerlin2013}.  In this case, the voltage dynamics are given by

\begin{equation*}
C_m \dot{\boldsymbol{V}} = -g_L (\boldsymbol{V}-E_L) - g C_m \Omega_f \boldsymbol{\rho} \ast h_r(t) + D\mathbf{c}(t),
\end{equation*}

\noindent where $-g_L (V-E_L)$ represents the leakage current. Each neuron receives synaptic input from other cells in the computing network as well as external input.  The external input is given by $D\mathbf{c}(t)$ where $D$ is a $N\times J$ matrix of input weights, and $\mathbf{c}(t)$ is the $J\times 1 $ vector of inputs introduced above. The synaptic input is given by $g C_m \Omega_f \boldsymbol{\rho} \ast h_r(t)$ where $\Omega_f$ is the network connectivity matrix, $g C_m $ scales the strength of the synaptic input, and $h_r(t)$ is a single exponential synapse (see Materials and Methods section \nameref{optimalnet} for details). 

Figure \ref{NetSchem1} (a) illustrates the resulting network structure in the simplest possible case.  This is a network consisting of a single neuron that receives stimulus input as well as input from recurrent (here, autaptic) connections, and a decoder $\mathbf{\hat{x}}$ that reads out the computation from the single neuron's spike train. Figure \ref{NetSchem1} (b) shows the resulting network behavior. For the examples in this figure, the network performs leaky integration on a single-variable, square wave input (i.e., the matrix $A$ is simply $-a_d$).  The upper plots show the decoded signal ($\hat{x}(t)$) from the spiking output of a single neuron (red traces) plotted against the actual desired signal $x(t)$ (dashed black lines) along with the neurons' voltage trace (lower panels).  In the first column, we illustrate the output of a single neuron from the leaky-integrate-and-fire (LIF) network of \cite{boerlin2013}.   Comparing the red decoded signal and the actual desired signal $x(t)$ demonstrates the principle of spike-based computation in action:  when the decoded signal deviates too far from the desired signal, an additional spike is triggered, and the process repeats.  
 
In the next column, we replace the exponential kernel used for decoding the network spike trains with a  double-exponential function (first arrow), as described above, which results in an LIF network without instantaneous ($\delta$-function) synaptic dynamics.  Next, as real neurons contain a variety of intrinsic currents, we replace the linear leakage current with generic Hodgkin-Huxley-type (HH) ionic currents:

\begin{equation*}
C_m\dot{\mathbf{V}} = \mathbf{I_{ion}}(\mathbf{V}) - g C_m\Omega_f \boldsymbol{\rho}\ast h_r(t) +  D\mathbf{c}(t),
\end{equation*}

\noindent where $\mathbf{I_{ion}}(\mathbf{V})$ represents the sum of all ionic currents and also depends on the corresponding dynamical gating variables. The third column in Figure \ref{NetSchem1} (b) illustrates how the network behaves with this change to the intrinsic voltage dynamics (labeled as adding ``spike currents'').

In general, the addition of such ionic currents to voltage dynamics will disrupt the ability of the network to accurately perform a given computation.  This is because the large excursions of the membrane potential during the action potential will cause the voltage of the individual neurons to deviate from their derived optimal relationship with the error. However, in Materials and Methods section \nameref{derivekernels}, we show that incorporating a new synaptic kernel in both the voltage and decoder dynamics allows the network to effectively compensate for the inclusion of ionic currents, so that it can perform the required computation with improved accuracy compared to the network where these compensation currents are not included. This new synaptic kernel, which we denote by $\tilde{\eta}(t)$, is constructed to counteract the total change in voltage that occurs during a spike. We provide details on how this kernel is derived as well as how it is obtained for our simulations in Materials and Methods sections \nameref{derivekernels} and \nameref{getkernels} and in Figure \ref{CompEx}.  The resulting voltage dynamics and decoder are:

\begin{eqnarray}
&&C_m\dot{\mathbf{V}} = \mathbf{I_{ion}}(\mathbf{V}) - g C_m\Omega_f \boldsymbol{\rho}\ast h_r(t) -C_m\Omega_c\boldsymbol{\rho}\ast \tilde{\eta}(t)+ D\mathbf{c}(t) \label{e.net11} \\
&&\mathbf{\hat{x}} = W \ast \boldsymbol{\rho}(t), \label{e.net12}
\end{eqnarray}

\noindent where $\Omega_c$ is the connectivity matrix for the compensating synaptic connections and $W(t)$ is the new decoding kernel (given in Materials and Methods section \nameref{derivekernels}). The final column of Figure \ref{NetSchem1} (b) shows how the addition of this compensation current affects the output of a single neuron. For the single neuron case, this adds large fluctuations in the decoder output.  Thus, compared with the original effects of adding the spike-generating currents, it appears that the compensation current can decrease accuracy.  However, our simulations show that this effect only occurs for very small (fewer than 4 neurons) networks.  For larger networks, compensation allows the network to perform the computation with a high degree of accuracy, as we will show.

To show how the framework generalizes to larger networks, we plot the output of an example network of $N=4$ neurons. For this network, we take $\Gamma_{1,2} = a$ while $\Gamma_{3,4}=-a$, where $a$ is a constant.  The output weights $\Gamma$ also determine the connectivity structure of the network. This particular choice of $\Gamma$ will lead to a network with all-to-all connectivity. The matrix $D$ that scales the stimulus input also depends on $\Gamma$: the network structure that allows the system to perform accurate spike-based computations requires that $D \propto \Gamma^T$ (see Materials and Methods sections \nameref{optimalnet} and \nameref{derivekernels}). This implies that neurons $1$ and $2$ ($3$ and $4$) will be depolarized (hyperpolarized) when $c(t)$ is positive. The cartoon in Figure \ref{NetSchem2} (a) shows the structure of this network.  

We next explore the output of our example 4-cell network. Here, the input to the network is a simple square-wave function of time, taking a fixed positive value from $100$ to $200$ $ms$ and a fixed negative value from $200$ to $300$ $ms$.  Figure \ref{NetSchem2} (b) shows the resulting spike rasters. The individual spike times are highly irregular, and the upper (lower) two cells appear to be more active when the input is positive (negative).  In Figure \ref{NetSchem2} (c), we again plot the network estimate $\hat{x}(t)$ (red) against the actual signal $x(t)$ (black dashed).  In addition, we also plot what the network estimate would be had the compensating synapses not been included (grey trace).  This shows that compensation indeed corrects for systematic biases.  Lastly, Figure \ref{NetSchem2} (d) plots the voltage trace for an example neuron.  There are two key points to take away from this final panel.  The first is that the synaptic input is not overwhelming the intrinsic spike-generating currents.  Indeed, one way to force the network to behave like an IF network would be to increase the synaptic gain so that the synaptic input is much larger than the intrinsic currents; this is clearly not the case here.  The second point to take away from the plot is that the membrane potentials and spike times of individual neurons appear highly irregular. 

The above examples, in implementing Equations~\eqref{e.net11}-\eqref{e.net12}, used a special choice for the matrix $A$ that defines the linear computation implemented by the network; here, we set $A=-a_d$ so that the connectivity matrix for the double-exponential function synapses is zero (see below and Materials and Methods section \nameref{derivekernels}).  For an arbitrary choice of $A$, the network dynamics are given by

\begin{eqnarray}
&&C_m\dot{\mathbf{V}} = \mathbf{I_{ion}}(\mathbf{V}) + gC_m\left[\Omega_s\boldsymbol{\rho}\ast\alpha(t)-\Omega_f\boldsymbol{\rho}\ast h_r(t)\right]-C_m\Omega_c\boldsymbol{\rho}\ast\tilde{\eta}(t)+D\mathbf{c}(t) +\sigma_V {\boldsymbol \xi}(t),
\label{e.fullnet}
\end{eqnarray}

\noindent where $\Omega_s$ represents the ``slow'' (compared to the exponential ``fast'' synapses) synaptic connectivity matrix.  This effectively corresponds to the decoded estimate $\hat{x}(t)$ being fed back into the network, which allows the network to perform more general computations on inputs. The parameter $g$ scales the strength of both the slow and fast synapses in the network.

Lastly, in Equation~\eqref{e.fullnet} we also added a white noise current ($\sigma_V {\boldsymbol \xi}(t)$), drawn independently for each cell, to our voltage evolution equations.  This represents random synaptic and channel fluctuations as well as noisy background inputs, but, as we will see below, also serves a functional role in decreasing network synchrony.


\subsection*{Homogeneous Integrating Network}

For the remainder of the paper, we focus on the case of a network of neurons with spike-generating currents based on the Miles-Traub model \cite{traubmiles1995,hoppensteadtpeskin2001} (Materials and Methods section \nameref{params}) which contains HH-type sodium, potassium, and leakage ionic currents. Although we use a specific model, similar results were obtained with different neuron models, e.g. a fast-spiking interneuron model \cite{erisir1999} and different sodium, potassium, and leakage current kinetic and biophysical parameters taken from \cite{mainen1995}. We will initially show how such a spiking network can integrate a one-dimensional stimulus input.  In terms of the notation previously introduced, this corresponds to the case where the number of inputs, or dimensionality, $J=1$ and the matrix $A=0$. We choose the input connections such that $\Gamma_k = a$ for half of the cells in the network, $k=1,...,N/2$, and $\Gamma_k = -a$ for the remaining half, $k=N/2+1,...,N$.  Thus, the network has all-to-all connectivity (recall that the network connectivity matrices depend on $\Gamma$, for example, $\Omega_f \sim \Gamma^T\Gamma$); the input to individual neurons within the ``first" or ``second" half of the network differs only via their (independent)  background noise terms.  With this configuration, half of the network will be depolarized when the stimulus input $c(t)$ is positive, while the other half will be hyperpolarized.  We will refer to the depolarized half as the ``stimulus-activated'' population and the hyperpolarized half as the ``stimulus-depressed'' population. Note that this distinction does not refer in any way to excitatory vs. inhibitory neurons, as in our formulation neurons can both excite and inhibit one another, a point that we will return to later.  The addition of voltage noise in this case is critical as the network is very homogenous and will synchronize in the absence of noise.  We systematically explore the dependence of network performance on the noise level (as well as other parameters) in a later section.

For purposes of illustration, the network was driven with two different types of inputs $c(t)$, a box function and a frozen random trace generated from an Ornstein-Uhlenbeck (OU) process \cite{uhlenbeck1930} (Figure \ref{NetEx}(a); see Methods for details). The remainder of Figure \ref{NetEx} shows the resulting output for a $400$-neuron network, integrating a box input in panels (a)-(e) and integrating the frozen random trace in panels (f)-(j).  Figures (a) and (f) plot the different inputs, while (b) and (g) show the raster plots for all $400$ neurons.  The neurons spike fairly sparsely and highly irregularly.  The network estimates, $\hat{x}(t)$ (red trace), along with the true signal $x(t)$ (blue trace) are shown in (c) and (h). The network is able to track both the box and OU inputs with a high degree of accuracy: the relative error ($||x-\hat{x}||_2/||x||_2$) between the estimate and the actual signal is $0.07$ for (c) and $0.07$ for (h). To illustrate the improvement in accuracy due to the synaptic inputs that compensate for spike-generating currents (see Materials and Methods section \nameref{derivekernels}), we also plot signal estimates from a network where this compensation was not included (grey traces).  For these estimates, the relative error is $0.60$ in (c) and $0.40$ in (h);  thus, our compensating synapses yield an almost tenfold increase in accuracy. 

Next, we show the population-averaged firing rates for the stimulus-activated (magenta) and stimulus-depressed populations (green) in (d) and (i).  Figure (d) shows that in the absence of input, the populations maintain persistent activity for roughly $500$ $ms$.  This is consistent with observations of neural activity during working memory tasks \cite{jonides2008}.  However in panel (i), the firing rates of the populations fluctuate depending upon the input.  Lastly, (e) and (j) plot the average autocorrelation functions for the spiking activity of neurons in the different populations.  These display a clear refractory effect, and small tendency to fire in the window that follows.  Differences between the stimulus-activated and stimulus-depressed populations, especially for the box function input, are likely due to the different firing rates and inputs that the two populations receive.  We explore these spiking statistics further in the section that follows.

\subsection*{Dynamics Underlying Network Computation}

We next show that our network displays two key features of cortical networks: the spike times of the network are irregular and Poisson-like, and there is a tight balance between excitation and inhibition for each neuron in the network.  Figure \ref{ISIFig} shows responses from the homogeneous integrator network introduced in the previous section with a box function input stimulus.  The irregularity of spike times is illustrated by the voltage trace of an example neuron in the network, in Figure \ref{ISIFig}(a).  To quantify this irregularity, we generated a histogram of the inter-spike intervals (ISI) during the period of zero input where the firing rates are nearly constant \ref{ISIFig}(b). To generate the histogram, we simulated the response of the network during $800$ repetitions of the box function input.  The only thing that varied between trials was the realization of the additive background noise current.  The ISIs follow an almost exponential distribution, see inset, and the coefficient of variation (CV) is $0.86$.  Thus, the spiking in our network is, by this measure, less variable but not far from what we would expect for Poisson spiking (which would yield a CV=1) or levels of variability that have been observed in cortical networks \cite{faisaletal2008,shadlen1998}.  

We also explore the trial-to-trial variability of individual neurons in the network.  Figure \ref{ISIFig}(c) shows a raster plot with the spike times of two example neurons over $20$ different trials.  The upper (lower) dots correspond to the spike times of a neuron from the stimulus-activated (stimulus-depressed) population.  One can see that the spike times of individual neurons vary considerably between trials.  To quantify this, we computed the time-averaged Fano factors for each neuron in the network (Materials and Methods section \nameref{Fano}). The Fano factor gives a measure of the trial-to-trial variability of individual neurons.  For the stimulus-activated population, the time averaged Fano factor, averaged across the population, is $0.515\pm 0.003$, while for the stimulus-depressed population, it is $0.761\pm 0.002$. For a time homogeneous Poisson process, one would expect a Fano factor of 1.  Thus, by this measure, neurons in both populations display variable spiking from trial-to-trial, but less variable than what would be expected from a Poisson process.    

By examining the total excitatory and inhibitory current that each neuron receives, we can check whether the network is in the balanced state \cite{haideretal2006,okunandlampl2008,vanvreeswijkandsompolinsky1996}.  To do this, we compute the total positive (negative) input a cell receives.  A complication here is that the $\tilde{\eta}(t)$ kernel changes sign; to deal with this, we rewrote the kernel as a difference of two separate, positive kernels, i.e., $\tilde{\eta}(t)=\tilde{\eta}_p(t)-\tilde{\eta}_n(t)$, and computed the resulting current from each kernel.  We also ignore the noisy background current for visualization purposes as similar results were obtained when the noise is included. Figure \ref{BalanceFig}(a) shows the total excitatory (red) and inhibitory (blue) current for an example neuron in the network.  Note that while the balance is imperfect (as shown by the inset), the two currents do appear to track each other fairly well.  Panel (b) shows the total excitatory (red) and inhibitory (blue) current averaged over all neurons in the network.  This shows that the currents are tightly balanced at the level of the entire network, which is typically what one finds when deriving so-called balanced networks \cite{brunel2000,limandgoldman2013,ostojic2014,vanvreeswijkandsompolinsky1996}

Next, we demonstrate that, even after altering the synaptic time scales and including spike-generating currents, neurons in the network still perform predictive coding by firing when their projected error signal is large. We computed the spike-triggered error (STE) for the network by aligning the projected error signal for each neuron $k$ ($\Gamma_k(x(t)-\hat{x}(t))$) to that neuron's spike times, averaging across all spike times and then averaging over all neurons (Materials and Methods section \nameref{STEcomp}), Fig. \ref{BalanceFig}(c).  The STE is indeed largest at the time of the spike and rapidly decreases right after the spike, indicating that spikes do in fact decrease the error.  The oscillatory behavior of the STE is indicative of the fact that there is some amount of synchrony in the spike times of the network.

Signatures of spike-based computation are also present in the subthreshold membrane potentials of cells in our network.  First, Figure \ref{VcorrFig} (a) shows the trial-averaged cross-correlogram (see Materials and Methods section \nameref{Vcorr}) between the subthreshold voltages of two example cells in the stimulus-depressed population (blue solid trace) and two example cells in different populations (red dashed trace).  The voltage traces of cells within the same population appear to be correlated over short time lags, as we expect from the fact that neurons in the same population receive highly similar synaptic input.  Meanwhile, voltages of cells in different populations are anti-correlated.  Thus, cells in different populations can be differentiated via correlations in their subthreshold voltages.  Next, we explore the voltage statistics of single cells. Figure \ref{VcorrFig} (b) shows the voltage power spectrum of an example cell in the stimulus-depressed population (solid trace).  For comparison, the power spectrum of an isolated neuron that only receives background noise input is shown in the dashed trace.  It appears that noise input drives the peak in the power spectrum around $40$ $Hz$, while the fast predictive coding implemented by the feed-forward input and lateral connections is responsible for the remaining peak around $150$ $Hz$ (Figure \ref{VcorrFig} (c) gives a closer view of this second peak.)  The presence of this second peak is therefore another prediction of the spike-based predictive coding framework.

\subsection*{Network Creates ``Good'' Correlations that Reduce Decoding Variance}

We now explore the structure of correlations that emerge among the spikes of different cells in the network, and whether these correlations are beneficial or harmful to the network's encoding of an input that has been integrated over time.  Specifically, we ask whether these coordinated spike times increase or decrease the variance of the decoded signal around its mean value.  As shown in Materials and Methods section \nameref{decvarsec}, the variance of the decoded signal is given by

\begin{equation*}
\textrm{var}(\hat{x})=\nu_{\hat{x}}=\Gamma C^{\alpha} \Gamma^T + \frac{1} {g^2}\Phi C^{\tilde{\eta}}\Phi^T -\frac{2} {g} \Gamma C^{\alpha \tilde{\eta}} \Phi^T,
\end{equation*}

\noindent where $C^{\alpha}_{ij}=\textrm{cov}(\rho^{\alpha}_i,\rho^{\alpha}_j)$, $C^{\tilde{\eta}}_{ij}=\textrm{cov}(\rho^{\tilde{\eta}}_i,\rho^{\tilde{\eta}}_j)$, and $C^{\alpha \tilde{\eta}}_{ij}=\textrm{cov}(\rho^{\alpha}_i,\rho^{\tilde{\eta}}_j)$ are the average covariance matrices of the spike trains convolved with the two synaptic kernels, i.e., $\boldsymbol{\rho}^Y = \boldsymbol{\rho}\ast Y(t)$, $Y\in\{\alpha, \tilde{\eta}\}$. This quantity measures the variability of the network estimate around its average value; lower values of this variance correspond to highly repeatable network estimates from one trial to the next. If the neurons in our network were independent, then the off-diagonal terms in these covariance matrices would all be zero.  Thus, the variance of an independent decoder $\nu^{\textrm{ind}}_{\hat{x}}$ would have the same form as the above equation, except that the off-diagonal terms of the covariance matrices would be set to zero.  The ratio $\nu^{\textrm{ind}}_{\hat{x}} /\nu_{\hat{x}}$ measures the reduction in decoding variance caused by the structure of pairwise interactions between neurons in the network.  The larger this ratio is, the greater the benefit of pairwise correlations between cells.  If the neurons in our network were indeed independent, then this ratio would be $1$.

How do correlations affect decoding variance in the homogeneous integrator network?  For both of the different inputs, the structure of pairwise interactions between neurons causes a roughly fivefold decrease in the variability of the network estimate: for the box function input, the reduction in decoding variance is $5.0$, while for the OU input, it is $5.8$.  To gain insight into how the correlation structure of the network causes this, Figure \ref{CorrFig} plots the population-averaged correlation coefficients and cross-correlograms for the homogeneous integrator network.  We first focus on the case of the box input function. In Figure \ref{CorrFig}(a) we show a histogram of the population-averaged pairwise correlation coefficients for both the stimulus-activated (magenta) and stimulus-depressed (green) populations.  Neurons in both populations appear to have weak (and slightly negative) pairwise interactions with one another on average: the mean correlation coefficient for the stimulus-activated (stimulus-depressed) population is $-1.3\times10^{-3}$ ($-0.3\times10^{-3}$). On the other hand, Figure \ref{CorrFig}(b) shows that the pairwise correlation coefficients between cells in the two different populations are small but positive, with a mean of $3.3\times 10^{-3}$.  Thus, the network reduces decoding variance by creating negative correlations between neurons that represent the same aspect of the stimulus, and positive correlations between neurons that represent different aspects of the stimulus.  From a coding perspective, these represent ``good'' correlations as the negative correlations between cells in the same population act to reduce redundancy, while the positive correlations across populations allow for some of the background noise to be cancelled out when the estimates from two populations are subtracted \cite{averbeck2006,hu2014}.  This can also be seen in the cross-correlograms of the different populations, Fig. \ref{CorrFig}(c) and (d). 

The situation is very similar for the OU stimulus input as shown in Figures \ref{CorrFig}(e)-(h).  There are slight differences in that the correlation coefficients are more broadly distributed, \ref{CorrFig}(e), and the correlation structure of the stimulus-activated and stimulus-depressed populations are more similar than for the box function stimulus.  This is likely due to the fact that, with the OU stimulus, the two populations receive a more similar range of inputs over time.

We have shown that the structure of pairwise interactions between neurons in the network acts to greatly reduce the variability of the network estimate of the underlying computation on a stimulus input.  This already reveals a difference between this framework and the underlying assumptions of a rate model, in which neurons in the network are assumed to be statistically independent.  As such, one could shuffle the spiking output of individual neurons from different trials and the rate-based computation would suffer no loss in accuracy.  However, for the predictive coding network, it was shown that the structure of interactions between spike trains for individual neurons from trial to trial is important to the accuracy of the desired computation \cite{boerlin2013}.  To give a more direct illustration of this effect with our current network, we explored how the relative error between the decoded network estimate and the actual signal varied as we replaced an increasing number of spike trains with variations recorded from separate trials ("shuffled" trains). 

Figure \ref{ReconErr} (a) plots the average relative error between desired ($x$) and network-decoded ($\hat x(t)$) signals (see Materials and Methods section \nameref{errormetrics}) as a function of the number of shuffled spike trains, for the box function input.  As expected, the error increases with the number of shuffled trains and reaches its maximum when all spike trains are taken from separate trials.  To see how the shuffling affects the network estimate, we show an example decoded estimate (red) plotted against the true signal (blue) in (b) when all spike trains are taken from the same trial.  In Figure \ref{ReconErr}(c), we plot the estimate decoded from entirely shuffled spike trains, where all are taken from different trials.  As also expected from the previous section, the effect of shuffling spike trains appears to increase the magnitude of the fluctuations of the decoded estimate around its mean value.  Figures \ref{ReconErr}(d)-(f) show that the situation is similar with the OU stimulus, although it is more difficult to see the effects on the decoded signal due to the fluctuations in the OU signal itself.

\subsection*{Sensitivity to Variation in Synaptic Strength and Noise Levels}\label{paramsens}

Our previous examples of the behavior of the homogeneous integrator system made use of a particular choice of network parameters.  We now explore the sensitivity of its performance to changes in these parameters.  In particular, we vary the strength of the fast and slow synaptic input, $g$, and the strength of the added voltage noise, $\sigma_V$.  For the homogeneous integrator network, these two parameters have the largest effect on performance as $g$ effectively scales the strength of synaptic connectivity between neurons in the network and $\sigma_V$ creates a level of heterogeneity in the individual voltage dynamics that prevents cells from synchronizing. We will show that the performance of our network is fairly robust to changes in these parameters.

We quantify network behavior using several measures. As before, the accuracy of the computation is evaluated using the relative error between the network estimate and the true signal. To assess the firing properties of the network, we compute a population synchrony index introduced by \cite{golomb2007} (Materials and Methods section \nameref{popsync}),  and the coefficient of variation of the interspike intervals during periods of zero stimulus input (for the box function input). We also track the maximum population-averaged firing rate, to ensure that the populations are not firing at unrealistically high levels.  Because similar results were obtained with the OU stimulus, we only report these metrics for the box function stimulus.

We first investigate how the level of population synchrony interacts with the accuracy of the network and neuronal firing rates. Figure \ref{PerfMetrics}(a) plots the population synchrony index as a function of the synaptic gain $g$ for three different values of the noise strength. The population synchrony has a $U$-shaped dependence on $g$; this is easiest to see at the smallest noise level.  When the population synchrony is high, the relative error is large (Figure \ref{PerfMetrics} (b)) and firing rates approach unrealistic levels (Figure \ref{PerfMetrics} (c)).  Thus, desynchronizing the firing dynamics of individual neurons in the network by increasing the noise to moderate levels improves network accuracy.  Our interpretation is that moderate noise distributes the computation more efficiently among individual neurons.  If the noise is too small, then individual neurons behave too similarly and eventually synchronize, effectively reducing the dimensionality of the network and also the computational power. When the noise is too large, the computation is overpowered by the noise.

Figure \ref{PerfMetrics} (b) plots the relative error between the network estimate and the true signal as a function of $g$ for three different noise levels.  As in Fig.~\ref{PerfMetrics} (a), for the first two noise levels (blue and magenta traces), the error appears to display an almost $U$-shaped dependence on $g$, indicating that there is an optimal choice for $g$ that minimizes the error for each noise level.  This value of $g$ also corresponds to the lowest value of the population synchrony index.  However, for the largest noise level (red trace), the error monotonically decreases as $g$ is increased. This could be indicative of the fact that, for this noise level, the population remains fairly desynchronized for a wide range of $g$ values. The effects of increasing the noise also depend on the value of $g$.  For small $g$, increasing the noise level first acts to decrease the error (compare blue to magenta), but then drives it to its highest level (red trace).  However, when $g$ is larger, noise appears to always cause the error to decrease.  For reference, the black circle on the magenta trace shows the values of $g$ and $\sigma_V$ that were used in the previous sections.

How do these parameter choices affect the networks' firing rates? Like the relative error traces in (b), the maximum population-averaged firing rates, Figure \ref{PerfMetrics} (c), also display a $U$-shaped dependence on $g$, and the shallowness of the $U$ increases as the noise level is increased.  This indicates that with increasing noise, there is a larger range of $g$ values that lead to low firing rates.  Lastly, Figure \ref{PerfMetrics} (d) plots the CV of the ISIs of the network during the period of zero stimulus input.  For moderate noise and moderate $g$, the network maintains CV on the order of $0.8$. 

In conclusion, network performance is not highly sensitive to changes in synaptic strength $g$ or to the level of added voltage noise, as there exist many combinations of choices that lead to similar network performance.

\subsection*{Recording from a Subset of Neurons}\label{sec:varyNrecorded}

Until now, we have assumed that the decoder has access to all neurons in the network that is performing the computation on the input; that is, we have fixed our network size at $N=400$ cells and have examined its performance using the spiking output of all $400$ cells.  However, when recording from real neural circuits, it is more likely that one would be measuring from a subset of cells involved in a given computation.  The same is possible for different circuits ``downstream'' of a computing network.  We explore how the reduction in decoding variance and the decoding error scales with the number of simultaneously recorded neurons.

Figure \ref{recsubsetfig} (a) plots the reduction in decoding variance $\nu^{\textrm{ind}}_{\hat{x}} /\nu_{\hat{x}}$ as a function of the number of simultaneously recorded neurons $M$ for the homogeneous integrator network with the box function input stimulus.  The simulated network size was fixed at $N=400$.  To compute the reduction in decoding variance for a smaller network of size $M$, a random subset of $M$ spike trains was chosen from a single simulated trial of the full network.  We then computed the necessary covariance matrices using these spike trains, and averaged these matrices over all $800$ trials.  These averaged covariance matrices were used to compute the ratio $\nu^{\textrm{ind}}_{\hat{x}} /\nu_{\hat{x}}$ according to the formulae given in Materials and Methods section \nameref{decvarsec}.  The solid trace in (a) plots the result of these numerical simulations whereas the dashed trace plots the approximation,

\begin{equation}
\frac{\nu^{\textrm{ind}}_{\hat{x}}} {\nu_{\hat{x}}} \approx \frac{1} {1-\frac{a_{\alpha}+d_{\alpha}} {2}+M\frac{a_{\alpha}+d_{\alpha}-2c_{\alpha}} {4}}\label{dcvarapprox},
\end{equation}

\noindent where $a_{\alpha}$ ($d_{\alpha}$) is the mean correlation coefficient between cells in the stimulus-activated (stimulus-depressed) population and $c_{\alpha}$ is the mean correlation coefficient between cells in the two different populations. Note that these correlation coefficients were computed using all $N$ cells in the simulated network. 

Figure \ref{recsubsetfig} (b) plots the square root of the decoding error, 

\begin{equation*}
\int_0^T[x(s)-\hat{x}(s)]^2ds,
\end{equation*}

\noindent as a function of the number of simultaneously recorded neurons.  As the number of recorded neurons increases, the decoding error initially decreases as $1/\sqrt{\textrm{Number of Recorded Neurons}}$ (black dashed line), similar to what one would expect for independent Poisson spiking, as implicitly assumed in many rate models.  However, as the number of recorded neurons is increased further, the error from the spiking network decreases faster than $1/\sqrt{\textrm{Number of Recorded Neurons}}$.

The predictions of our network about how the reduction in decoding variance and the decoding error both scale with the number of simultaneously recorded neurons could in principle be tested with dense multi-electrode arrays or optical imaging.  However, these predictions would have to be modified to incorporate the effects of shared sensory noise or noise in the output of the decoder.

\subsection*{Varying Network Size}\label{sec:varyNsize}

We now explore how the total number of neurons in the network, $N$, affects the fidelity of the computation.  We limit our analysis to integration of the box function stimulus.  As derived in Materials and Methods section \nameref{VaryNscaling}, we scale both the entries of the matrix $\Gamma$ and the synaptic gain parameter $g$ with $1/N$.  Using this scaling allows the total input to each neuron in the network to remain constant as the network size is varied.  

Figure \ref{VaryNFig} shows the results of these simulations.  In particular, we explore how the population synchrony index, the relative error, the time- and population- averaged firing rate, and the integrated error vary as the network size is increased. In all plots, the cyan trace at $N=400$ corresponds to the parameters used in our previous network simulations. Panel (a) plots the inverse of the synchrony index as a function of $N$ for $4$ different values of the parameter $c_0$, which scales the synaptic gain (that is, $g \sim c_0 \frac{1} {N}$).  This highlights the differences between the curves corresponding to the different values of $c_0$.  It is clear that synchrony tends to always decrease as the network size is increased, though the maximum level of synchrony reached as well as the rate at which it decreases with N are both affected by $c_0$.  Thus, as we have seen previously in Figure \ref{PerfMetrics}, increasing the synaptic gain can lead to increased population synchrony (compare the cyan and magenta traces).  Figure (b) plots the relative error as a function of $N$.  For small values of $c_0$, the error initially increases with $N$, but quickly reaches an asymptote and remains constant with further increases in network size (blue trace).  As $c_0$ is increased, we quickly see a transition in the curves as the relative error now begins to decrease with $N$.  Increasing $c_0$ initially causes the error to drop off faster with $N$ (compare the red and cyan traces), but too large of a value for $c_0$ cause the error to drop off more slowly with $N$ (compare the cyan and magenta traces).  Figure (c) plots the inverse time- and population-averaged firing rate during the period of zero stimulus input as a function of $N$.  As with the population synchrony index, the firing rates tend to decrease as $N$ is increased. 

In sum, Figures~\ref{VaryNFig} (a)-(c) illustrate that the computational error produced by the network, as well as its firing rates and synchrony, all tend to decrease for larger networks.  We next compare the trend in error against what would be naively expected in a simple ``rate network" -- that is, one in which each neuron fires according to a prescribed firing rate in a population, and does so with independent Poisson statistics.  In this case, we expect that the square root of the mean integrated squared error will scale like $1/\sqrt{N}$.  To compare the error in our spiking network, we plot the square root of the mean integrated squared error as a function of $N$ in Figure~\ref{VaryNFig} (d).   For $c_0=0.4$ (cyan trace), the error decreases as $1/\sqrt{N}$ (black dashed line) just as for the Poisson rate network.  However, for such rate networks, the firing rates of individual units are fixed and do not vary with the network size.  In our network, we clearly see a dramatic decrease in firing rates as network size grows, up to around $N=400$ cells (Fig. \ref{VaryNFig} (c)).  Further increases in network size past this point lead to minimal decreases in the firing rates. The fact that the firing rates for our network change with network size is a strong difference from a Poisson rate network.  Thus, even though our network produces a similar error scaling of error with $N$ as predicted under basic assumptions for firing rate networks, for network sizes between $100$-$400$ neurons,  it manages to do so in a more efficient manner -- it produces the same error with a lower average firing rate (i.e., fewer spikes).   

\subsection*{Beyond Pure Integration: Leaky Integration and Damped Harmonic Oscillations}\label{morecomp}

In this section, we highlight the generality of our approach by showing the output of the spike-based predictive coding networks that are performing computations other than ``pure" integration of its inputs over time. First, we study leaky integration, obtained in equation (\ref{xeqn}) by choosing $A$ to be negative (and continuing to take $J=1$ dimension for the signal $x(t)$). Figure \ref{LeakyIntFig} shows an example of the network performing leaky integration ($A=-10$) on the same box function input from Figure \ref{NetEx} (a). All other network parameters are the same as in Figure \ref{NetEx}.  The raster plot in Figure \ref{LeakyIntFig} (a) shows that the network still displays sparse irregular spiking when performing leaky integration. Figure \ref{LeakyIntFig} (b) shows the network estimate (red) plotted along with the actual signal (blue), demonstrating that the leaky integration computation is performed with a high degree of accuracy (the relative error is $0.08$).  Lastly, Figure \ref{LeakyIntFig} (c) shows the firing rates of both the stimulus-activated and stimulus-depressed populations; note that these eventually return to their baseline (pre-input) levels because the computation is leaky.

Next, we consider the computation of processing inputs through a two dimensional dynamical system that displays damped harmonic oscillations.  Here, the matrix $A$ is chosen as 

\begin{equation*}
A=
\begin{bmatrix}
\mu & -\omega \\
\omega & \mu
\end{bmatrix}.
\end{equation*}

\noindent In this case, the eigenvalues of the matrix $A$ are $\mu\pm i \omega$, and the solutions $\mathbf{x}(t)$ of the linear system (\ref{xeqn}) will display damped oscillations as long as $\mu<0$ (we use $\mu=-5$ and $\omega=20$).  We take $\Gamma$ to be a $2\times N$ matrix whose elements are chosen randomly. Lastly, we use as our input $\mathbf{c}(t)$ a vector with $c_1(t)$ being the box function input from Figure \ref{NetEx} (a) and $c_2(t)=0$. 

Figure \ref{DampedOscFig} plots the resulting network behavior.  We again see sparse irregular spiking with firing rates that eventually return to their baseline level (Figure \ref{DampedOscFig} (a) and (b)).  As the signal $\mathbf{x}(t)$ is two dimensional, the network estimate $\mathbf{\hat{x}}(t)$ is also two dimensional, and we plot both of the network estimates (red) along with the actual signals (blue) in Figure \ref{DampedOscFig} (c) and (d).  Once again, the network is able to perform the required computation with a high degree of accuracy (the relative error in (c) is $0.14$ and in (d) is $0.12$).

%
\section*{Discussion}

\subsection*{Synaptic Kinetics that Support Spike-based Computation}

We have shown that networks of neurons with voltage-dependent spike-generating currents and realistic synaptic timescales can perform accurate spike-based computations. These networks are derived based upon the premise that the voltage traces of individual neurons represent an error signal between the network estimate and the actual signal, and that spikes occur whenever the error becomes too large.  The key innovation we present that allows the network to accurately perform these computations is the inclusion of synapses with appropriate kinetics.  Two factors determine these kinetics.  We begin by assuming that signals are ``decoded" from the network with synapses that have finite timescales of rise and decay (``double-exponential" synapses).  Next, we account for the nonlinear dynamics of spike generating currents with ``compensating'' synapses, which allow the system to represent the projected error signal in the voltage traces of individual neurons.  It is important to note two limitations of these additional factors.  The first is that if the rate of rise of the double-exponential synapse is slower than the rate of change of the signal, then the accuracy of the computation will be affected.  This is the case because the network simply cannot respond quickly enough to accurately track the signal.  Secondly, the ``compensating'' synapses we introduce were designed to compensate for currents acting on the timescale of a single action potential.  Thus, slow adaptation currents are not accounted for by our approach. This will make it difficult for the network to maintain the persistent activity that is required when the desired computation is pure integration. Interestingly, however, we find that when the desired computation is leaky, simulations suggest that adaptation may have a minimal effect on the performance of the network. Thus, a network with strong adaptive currents is perhaps better suited to implement leaky integration or fast dynamics rather than perfect integration.

Our results prove the principle that mechanisms of spike-based computation previously derived for networks of idealized neurons and synapses \cite{boerlindeneve2011,boerlin2013} can be extended to settings closer to the underlying biophysics.  However, there is still distance to travel before we arrive at a ``realistic" biologically based system.  The compensating synapses are somewhat complicated functions of time.  Moreover, these and other synaptic connections provide both positive and negative currents following a spike, a clear violation of Dale's rule (although recently neurons that release both GABA and glutamate have been found in rodents \cite{root2014}).  More complex synaptic waveforms, and ones that change sign, could be implemented via intermediate synapses with different kinetics (for example, a pathway with delayed feedforward inhibition will produce first positive, then negative, synaptic current).   Furthermore, recent advances in learning temporal connections between neurons \cite{kennedy2014}, together with learning algorithms for the present spike-based computation framework \cite{bourdoukanetal2012},  provide a basis to potentially derive a learning rule for the compensation filters.  However, a question for future work is whether there are other network configurations that perform spike-based computation without the need for intermediate connections (as for simpler settings in \cite{boerlin2013}), and additionally with simpler synaptic waveforms than the compensating ones derived here.

\subsection*{Computing with Spikes and Computing with Rates}

As in \cite{boerlindeneve2011,boerlin2013}, our network approaches the notion of computations in neural circuits from the standpoint that a computation is distributed among the spike times of individual neurons.  This stands in contrast to many studies in which the computation is assumed to be carried at the level of averaged firing rates \cite{brunel2000,compteetal2000,goldmanetal2003,ostojic2014,renartetal2004,wongandwang2006}, and to other related studies that derive network dynamics that minimize squared error in signal representation~\cite{rozell08}. Demonstrating the importance of spike timing in our network, we compared the accuracy of the underlying computation before and after shuffling these times but preserving trial-averaged firing rates (Fig. \ref{ReconErr}).  Shuffling indeed reduced the accuracy, a fact we related to the structure of spike-time correlations produced by the network.   

What are the advantages of using such a precise temporal representation? It could be that distributing a computation among the spike times of individual cells endows the network with robustness to perturbations such as synaptic failure and lesions (as was demonstrated by \cite{boerlin2013}).  Moreover, a computation performed on the level of spike times opens the possibility that the underlying network structure could be learned via spike-based plasticity rules, as suggested by recent work \cite{bourdoukanetal2012}.   

Finally,  a precise temporal code may leverage the computational power of individual spiking neurons in a more efficient manner than rate-based approaches.  Traditional spiking network implementations of rate-based networks employ large amounts of added voltage noise to avoid synchrony, and large cell populations, so that the resulting population output is well described by ``mean field" rate equations \cite{brunel2000,compteetal2000,machensetal2005,ostojic2014,renartetal2004,wongandwang2006}.  Thus large populations of cells with noise-driven spiking represent signals in traditional rate-based approaches.  The possibility that spike-based computation might give rise to a significantly lower total error for a given population size.  This seems likely, given the results in \cite{boerlin2013} and in Results section \nameref{sec:varyNrecorded}. That is, the error in our networks can decrease faster than expected for a population of cells with independent spike times.  Making a more direct comparison to rate-based networks is an interesting area for future work.

This said, this spike-based approach to computation is not immune to problems with synchrony, and the need for additive noise to combat it.  We showed that there is an optimal level of this noise at which the network retains characteristics of spike-based computation.  Moreover, in the current work, we have used a very homogeneous population in which all neurons have the same spike-generating currents and magnitude of synaptic connectivity. Preliminary simulations suggest that more heterogeneous networks better avoid synchrony, and hence may be able to perform computations with higher accuracy.

\subsection*{Balanced Networks and Irregular Spiking Behavior}

Cortical neurons are known to display irregular Poisson-like firing \cite{shadlen1998,Sof+93}.  What might be the basis of the observed irregularity? Various authors have proposed that the variability is a result of a tight balance between the total excitatory and inhibitory current each neuron in the network receives \cite{renartetal2010,vanvreeswijkandsompolinsky1996}.  Indeed cortical networks have been shown to display such a balance between excitation and inhibition \cite{haideretal2006,okunandlampl2008,wehrandzador2003}.  

While successful in reproducing the Poisson-like firing of cortical neurons, it is only very recently been shown that balanced networks can be used to perform particular computations, including integration of inputs over time \cite{boerlin2013,limandgoldman2013}.  Our work contributes further results in this direction.  In contrast to much previous modeling work on balanced networks, the specific condition for balancing excitation and inhibition is not built into the derivation of spike-based computation.  Rather, the balanced state arises naturally as a consequence of optimizing the computation at the level of single cells --- that is, assuming that neurons represent a projected error signal in their voltage traces and spike when this error becomes too large \cite{boerlin2013}.  Furthermore, our network maintains this balanced state with a relatively small number of cells (e.g., Fig. \ref{NetEx}).  Thus, our work suggests that a computational unit in the brain may require dramatically fewer neurons than predicted by rate-based approaches  \cite{brunel2000,limandgoldman2013,ostojic2014}.  

Finally, nearly all modeling work on balanced networks makes use of simplified integrate-and-fire neuronal spiking dynamics.  Thus, it has remained unclear whether or not the balanced state can be maintained by a network of neurons with more complex spike-generating dynamics. We have shown here that it is indeed possible for a network of such neurons to display a tight balance between excitation and inhibition, and thus display irregular spiking.  

Even during a single trial, the neurons in our network display variable spiking behavior.  This can be seen from the population firing rates in Figure \ref{NetEx} (d). Even though the network maintains a constant decoded signal, the average firing rates fluctuate, indicating that not all neurons are displaying the same firing rate.  This phenomenon of variable neuronal activity underlying stable network stimulus representation is known to occur \cite{buzsaki2004,haider2009,tchumatchenko2011} and has recently been shown in a rate model network with a specific architecture \cite{druckmann2012}. Here, we show that stable stimulus representation with variable neuronal responses arises as a natural feature of networks that perform spike-based computation.

\subsection*{Sensitivity and Tuning}

The performance of network models that integrate inputs over time is typically quite sensitive to the choice of connection weights between neural populations \cite{seungetal2000}.  If the recurrent connections are either too strong or too weak, the activity of the network can either quickly increase to saturation or decrease to a baseline level.  Recent work by Lim and Goldman \cite{limandgoldman2013,limandgoldman2014} has shown that this sensitivity issue can be resolved in a rate-based network where inhibition and excitation are balanced. In particular, Lim and Goldman show that a balanced rate-based network of leaky-integrate-and-fire (LIF) neurons can robustly maintain information for working memory with irregular spiking.  Further work will be needed to assess whether the spike-based networks derived here have similar robustness to changes in network connection strengths; our preliminary studies suggest that they may not, as perturbing the network structure away from the optimally derived connectivity will lead to decreased network accuracy.  However, there is evidence to suggest that the optimal connectivity structure could potentially be learned, and maintained, by plasticity rules \cite{bourdoukanetal2012,kennedy2014}.



\newpage

\section{Legends}

\begin{figure}[h!]
\caption{{\bf Leaky integration with a single biophysical neuron.}  (a)  Cartoon illustrating the connectivity of a ``network" consisting of a single neuron ($N=1$).  The cartoon shows that the neuron receives stimulus input as well as input from synaptic connections to itself, and the decoder $\hat{x}(t)$ ``reads-out'' the computation from the spike trains of the network. (b)  Schematic of how the network is derived in the case of a single neuron.  The upper plots show the decoded signal ($\hat{x}(t)$) (red traces) plotted against the actual signal $x(t)$ (dashed black lines) along with the neurons' voltage trace (lower panels).  For the examples in this figure, the network is performing leaky integration on a box function input. In the first column, we illustrate the output of a single neuron from the LIF framework of Boerlin et al.  In the second column, we alter how the stimulus information is read-out from the spike-times of the network (first arrow) which results in an LIF network without instantaneous ($\delta$-function) synaptic dynamics. Going from the second to third columns, we add spike-generating, Hodgkin-Huxley-type ionic currents to the voltage dynamics.  The fourth column illustrates how the addition of the compensating synaptic kernel affects the output of the decoder.}\label{NetSchem1}
\end{figure}


\begin{figure}[h!]
\caption{{\bf Obtaining the compensating synaptic kernels.} The compensation kernel $\eta(t)$ was obtained by stimulating a single model neuron with a fluctuation (Ornstein-Uhlenbeck) current and keeping track of the times $t_j$ that the voltage crossed a threshold from below (black dashed line in first panel).  For each spike, we then obtain an action potential waveform $V_{AP}^j(t)$ for $t_j\leq t < t_j+t_s$, where $t_s$ will set the width of the $\eta(t)$ kernel (we take $t_s=4$ $ms$).  We then sum these traces to obtain the average waveform of the action potential $V_{AP}(t)$ (black dashed line in the second panel).   The kernel $\eta(t)$ is then the temporal derivative of this averaged action potential waveform (third panel). $\eta(t)$ represents an approximation to the total change in voltage of the neuron during an action potential.  Lastly,  $\eta(t)$ is convolved with an exponential function to obtain the synaptic kernel $\tilde{\eta}(t)$  (last panel). Note that $\tilde{\eta}(t)$ changes sign but also very rapidly goes to zero as time goes on.  }\label{CompEx}
\end{figure}


\begin{figure}[h!]
\caption{{\bf Leaky integration with a network of biophysical neurons.}  (a)  Cartoon of an example network of $N=4$ cells performing leaky integration. In the network, the upper two cells (magenta) are excited by positive stimulus input (indicated by the red lines) while the bottom two cells (green) are depressed (indicated by the blue lines).  Each neuron receieves the stimulus input as well as synaptic input from every other neuron in the network.  The spike trains of all four neurons are used in generating the network estimate $\hat{x}(t)$.  (b) Raster plot from the example network of four neurons. The input to the network in this case is a simple box function, with a fixed positive value from $100$ to $200$ $ms$ and a fixed negative value from $200$ to $300$ $ms$. (c) Network estimate $\hat{x}(t)$ (red trace) plotted against the actual signal $x(t)$ (black dashed trace). The grey trace shows the estimate obtained if the compensating kernel was not included in the network dynamics.  (d)  Voltage trace for the topmost neuron in the example network (top row of the raster plot in (c)).}\label{NetSchem2}
\end{figure}


\begin{figure}[h!]
\caption{{\bf Homogeneous integrator network.} We show the output of a network of $N=400$ cells where $\Gamma_k=0.1$ for $k=1,...,N/2$ (stimulus activated population) and $\Gamma_k=-0.1$ for $k=N/2+1,...,N$ (stimulus depressed population), $g=0.4$ $mV$, { $\sigma_V=0.08$ $\mu A/cm^2 \cdot \sqrt{s}$}, and $c_0=g$. All other parameters are given in Materials and Methods section \nameref{params}. Panels (a)-(e) show the output of the network with the box function input stimulus while (f)-(j) show the output with the OU stimulus input.  (a) and (f) plot the stimulus input  $c(t)$ into the network.  (b) and (g) show the raster plots of the $400$ cells in the network.  The top $200$ rows are the stimulus activated population while the bottom $200$ rows are the stimulus depressed population. (c) and (h) plot the network estimate (red) against the actual signal (blue) along with the estimate obtained if compensation was not included (grey). (d) and (i) plot the average firing rates for the stimulus activated (magenta) and stimulus depressed (green) populations.  (e) and (j) show the population averaged autocorrelograms.  }\label{NetEx}
\end{figure}


\begin{figure}[h!]
\caption{{\bf Neurons in the network display irregular spiking.} (a) Example voltage trace from a single neuron from the homogeneous integrator network with the box function input as in Figure \ref{NetEx}. The dashed line represents the threshold used for spike detection. (b) Histogram of the interspike intervals of the network during the period of zero stimulus.  The inset shows the same data re-plotted with the y-axis on a log scale.  The coefficient of variation in this case is $0.86$. (c) Raster plot of the spike times of two example neurons (one from each population) on $20$ different simulated trials.  The magenta (green) dots correspond to the spike times of a neuron from the stimulus activated (stimulus depressed) population.  To quantify the trial-to-trial variability, the time averaged Fano factor was computed for each neuron in the network, and then averaged over all cells in each population.  This gave $0.515 \pm 0.003$ for the stimulus activated population and $0.761 \pm 0.002$ for the stimulus depressed population.}\label{ISIFig}
\end{figure}


\begin{figure}[h!]
\caption{{\bf Neurons in the network display a tight balance between excitation and inhibition, and spike only when the error between the estimated and actual signal is large.} (a) Total excitatory (red) and inhibitory (blue dashed) currents (ignoring background noise) into an example neuron from the homogeneous integrator network of Figure \ref{NetEx}.  The inset shows a blow-up of a particular time period to show that currents track each other fairly well.(b)  Total excitatory (red) and inhibitory (blue dashed) currents averaged over all $400$ neurons in the network. The inset shows that, on average, the currents are nearly identical, and thus balanced. (c)  Average projected error signal aligned to the spike times of each neuron in the network, i.e., the spike-triggered error signal (see Materials and Method section \nameref{STEcomp}).  The error is largest around the time of a spike indicating that, on average, neurons spike when this projected error signal is large.  }\label{BalanceFig}
\end{figure}


\begin{figure}[h!]
\caption{{\bf Error signal affects the correlation of the subthreshold voltage activity in the homogeneous integrator network.} (a) Trial averaged cross-correlation between the subthreshold voltage activity of two cells in the stimulus-depressed population (blue solid trace) and two cells in different populations (red dashed trace). Cells in the same population (different populations) show correlated (anti-correlated) voltage activity over short time lags. (b) Trial averaged voltage power spectrum for an example neuron in the stimulus-depressed population (blue solid trace) and for an isolated cell with only background noise input (dashed trace). (c) Change in power (expressed in decibels) that occurs when synaptic connections are included (logarithm base $10$ of the solid trace in (b) divided by the dashed trace). Recurrent inputs contribute to the peak in power around $150$ $Hz$.  }\label{VcorrFig}
\end{figure}


\begin{figure}[h!]
\caption{{\bf Structure of spike time correlations for the homogeneous integrator network.} (a)-(d) show the  structure of spike time correlations for the network with the box function input while (e)-(h) show the structure of the network with the OU input.  (a) and (e) plot a histogram of the population averaged pairwise correlation coefficient between cells in the stimulus activated population (magenta) and between cells in the stimulus depressed population (green).  In (a) the mean correlation coefficient across trials for the stimulus activated (stimulus depressed) population is $-1.3\times 10^{-3}$ ($-0.3\times 10^{-3}$), while in (e) it is  is $-1.0\times 10^{-3}$ ($-0.7\times 10^{-3}$). (b) and (f) plot a histogram of the average correlation coefficient between cells in the two different populations.  In (b), the mean correlation coefficient across trials is $3.3\times 10^{-3}$, while in (f) it is $5.0\times 10^{-3}$. (c) and (g) plot the population and trial averaged  shift-predictor-corrected cross correlograms for the raw spike trains of neurons within the stimulus activated (magenta) and stimulus drepressed (green) populations.  (d) and (h) plot the population and trial averaged  shift-predictor-corrected cross correlograms for the raw spike trains of neurons in the two different populations.   }\label{CorrFig}
\end{figure}


\begin{figure}[h!]
\caption{{\bf Shuffling spike trains across trials distinguishes the network from a rate model.} We explore how the decoding error varies as we decode the spiking output of the network where we replace an increasing number of individual neuron spike trains with those from different trial simulations. The parameters are the same as those used in Figure \ref{NetEx}.  (a) plots the relative error $||x - \hat{x}||_2/||x||_2$ as a function of the number of replaced spike trains for the network with the box function input.  As the number of replaced spike trains is incresed, so does the error.  (b) plots an example network estimate (red) against the actual signal (blue) when no spike trains have been replaced.  (c) plots the network estimate (red) against the actual signal (blue) when all $400$ spike trains are taken from separate trials.  Notice how replacing spike trains increases the variability of the estimate around its mean.  For comparison, the relative error in (b) is $0.05$ while in (c) it is $0.09$.  (d)-(f) are the same as (a)-(c) except that the OU stimulus is used.  The relative error in (e) is $0.09$ while in (f) it is $0.18$. }\label{ReconErr}
\end{figure}


\begin{figure}[h!]
\caption{{\bf Dependence of network statistics on noise and synaptic gain parameters.} We explore how the network output changes as we vary the synaptic gain parameter $g$ and the strength of the voltage noise $\sigma_V$.  We set the parameter $c_0$ which scales the strength of the input to $c_0=g$ for every value of $g$ used. Because the results were similar for the frozen noise (OU) case, we only plot the results for the network with the box function input.  In panels (b-d) we indicate the parameter values used in the previous figures with a black circle. Error bars represent standard deviations over $300$ trials. (a) Population synchrony index (see Materials and Methods section \nameref{popsync}) as a function of $g$ for three different noise levels { $\sigma_V=0.04$ $\mu A/cm^2 \cdot \sqrt{s}$} (blue trace),  { $\sigma_V=0.08$ $\mu A/cm^2 \cdot \sqrt{s}$} (magenta trace), and { $\sigma_V=0.12$ $\mu A/cm^2 \cdot \sqrt{s}$} (red trace).  (b) Relative error between the estimate and the actual signal as a function of $g$.  (c) Maximum population averaged firing rate as a function of $g$.  (d)  Coefficient of variation of the ISIs during the period of zero stimulus input as a function of $g$.   }\label{PerfMetrics}
\end{figure}


\begin{figure}[h!]
\caption{{\bf Reduction in decoding variance and error scale with the number of recorded neurons.} We explore how the network output varies if we only have access to a subset of neurons in the full simulated network.  The parameters are the same as in Figure \ref{NetEx} and we only show results for the box function input. (a) plots the reduction in decoding variance (see Materials and Methods section \nameref{decvarsec}) as a function of the number of simultaneously recorded neurons $M$.  The solid trace shows the results from the numerical simulations while the dashed trace plots the analytical approximation that uses the mean correlation coefficients of full $N=400$ network.  (b) plots the square root of the integrated squared error between the estimate and the actual signal as a function of the number of simultaneously recorded neurons on a log-log plot.  The dashed trace is the line $1/\sqrt{\textrm{Number of Recorded Neurons}}$ which would be the prediction of a network of independent Poisson processes.  Notice that the error in our network initially decreases like $1/M$, but eventually begins to decrease at a much faster rate. } \label{recsubsetfig}
\end{figure}


\begin{figure}[h!]
\caption{{\bf Varying the simulated network size.} We explore how the network output varies when we change the total number of simulated neurons. We again use the homogeneous network with $\Gamma_k=a$ for $k=1,2,...,N/2$ and  $\Gamma_k=-a$ for $k=N/2+1,...,N$, { $\sigma_V=0.08$ $\mu A/cm^2 \cdot \sqrt{s}$}, and the box function input. As derived in the Material and Methods section \nameref{VaryNscaling}, we use the scaling $a=\frac{400} {N}$, $g=c_0\frac{40} {N}$, and the different colored lines correspond to different values for $c_0$. Error bars represent standard deviations over $900$ repeated trials. In (a) we plot the inverse of the population synchrony index as a function of the simulated network size for different values of the parameter $c_0$ which scales the gain of the synaptic input.  (b) relative error between the estimate and the actual signal as a function of $N$.  (c) the inverse of the time and population averaged firing rate during the period of zero stimulus input as a function of $N$.  (d)  Square root of the mean integrated error as a function of $N$ on a log-log plot.  The two dashed black lines plot $1/\sqrt{N}$ starting from the first cyan data point and the first magenta data point. } \label{VaryNFig}
\end{figure}


\begin{figure}[h!]
\caption{{\bf Leaky integration in the homogeneous network.} We show the results of the network performing leaky integration on the box function input. All parameters are the same as in Figure \ref{NetEx} except that $A=-10$; $c(t)$ is the same box function from Figure \ref{NetEx} (a). (a) shows the raster plot of the 400 cells in the network. The top 200 rows are the stimulus activated population while the bottom 200 rows are the stimulus depressed population. (b) plots the network estimate (red) against the actual signal (blue). (c) plots the average firing rates for the stimulus activated (magenta) and stimulus depressed (green) populations. }\label{LeakyIntFig}
\end{figure}


\begin{figure}[h!]
\caption{{\bf Damped harmonic oscillations.} We show the results of the network processing the box function input via damped harmonic oscillations.  All parameters are the same as in Figure \ref{NetEx} except that $\Gamma$ is now a $2\times N$ matrix and $\mathbf{c}(t)$ is a vector with $c_1(t)$ being the same box function from Figure \ref{NetEx} (a) and $c_2(t)=0$.  The elements of $\Gamma$ are chosen randomly with $\Gamma_{ij} \in \textrm{Unif}(0.002,1)$ for $i=1,2$ and $j=1,2,...,N/2$ and $\Gamma_{ij} \in \textrm{Unif}(-1,-0.002)$ for $i=1,2$ and $j=N/2+1,...,N$.  Also, the columns of $\Gamma$ are normalized so that $||\Gamma||_j=\sqrt{\Gamma_{1j}^2+\Gamma_{2j}^2}=0.06$. (a) shows the raster plot of the 400 cells in the network. The top 200 rows are the stimulus activated population while the bottom 200 rows are the stimulus depressed population. (b) plots the average firing rates for the stimulus activated (magenta) and stimulus depressed (green) populations. (c) and (d) plot the network estimates (red) against the actual signals (blue). }\label{DampedOscFig}
\end{figure}


\clearpage
\newpage

\section{Figures}

\begin{center}
\includegraphics[scale=1]{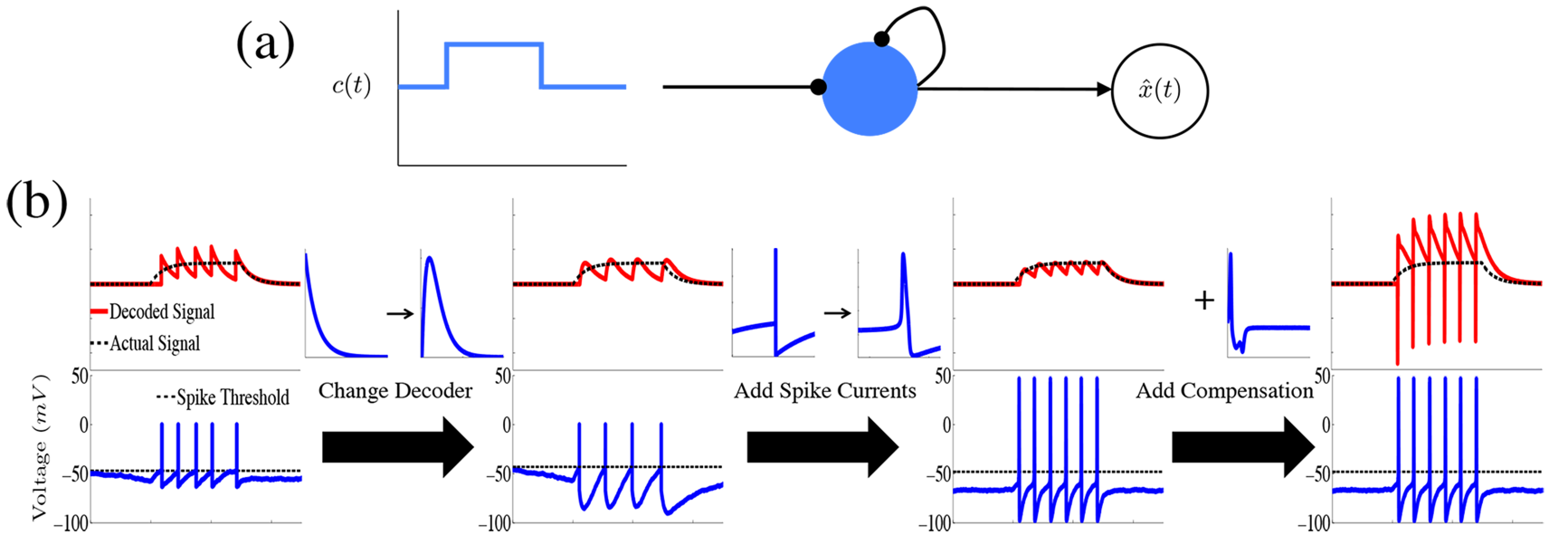}\\
{\bf Figure 1}
\end{center}


\clearpage
\newpage

\begin{center}
\includegraphics[scale=1]{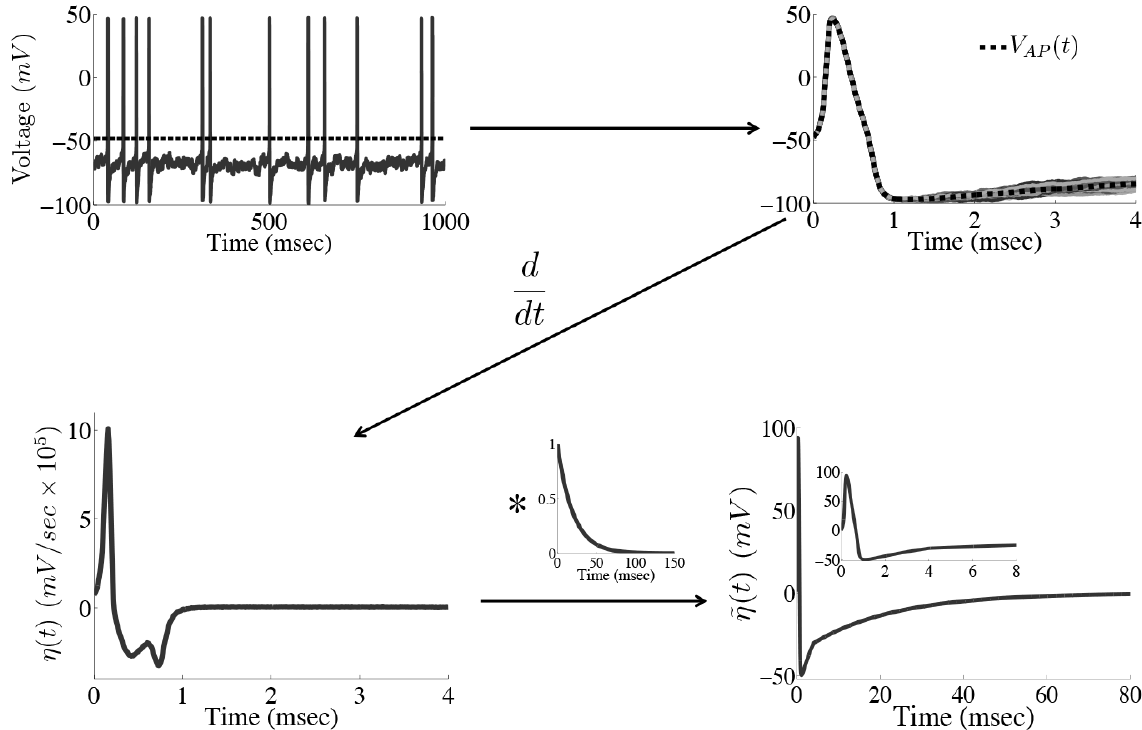}\\
{\bf Figure 2}
\end{center}


\clearpage
\newpage

\begin{center}
\includegraphics[scale=1]{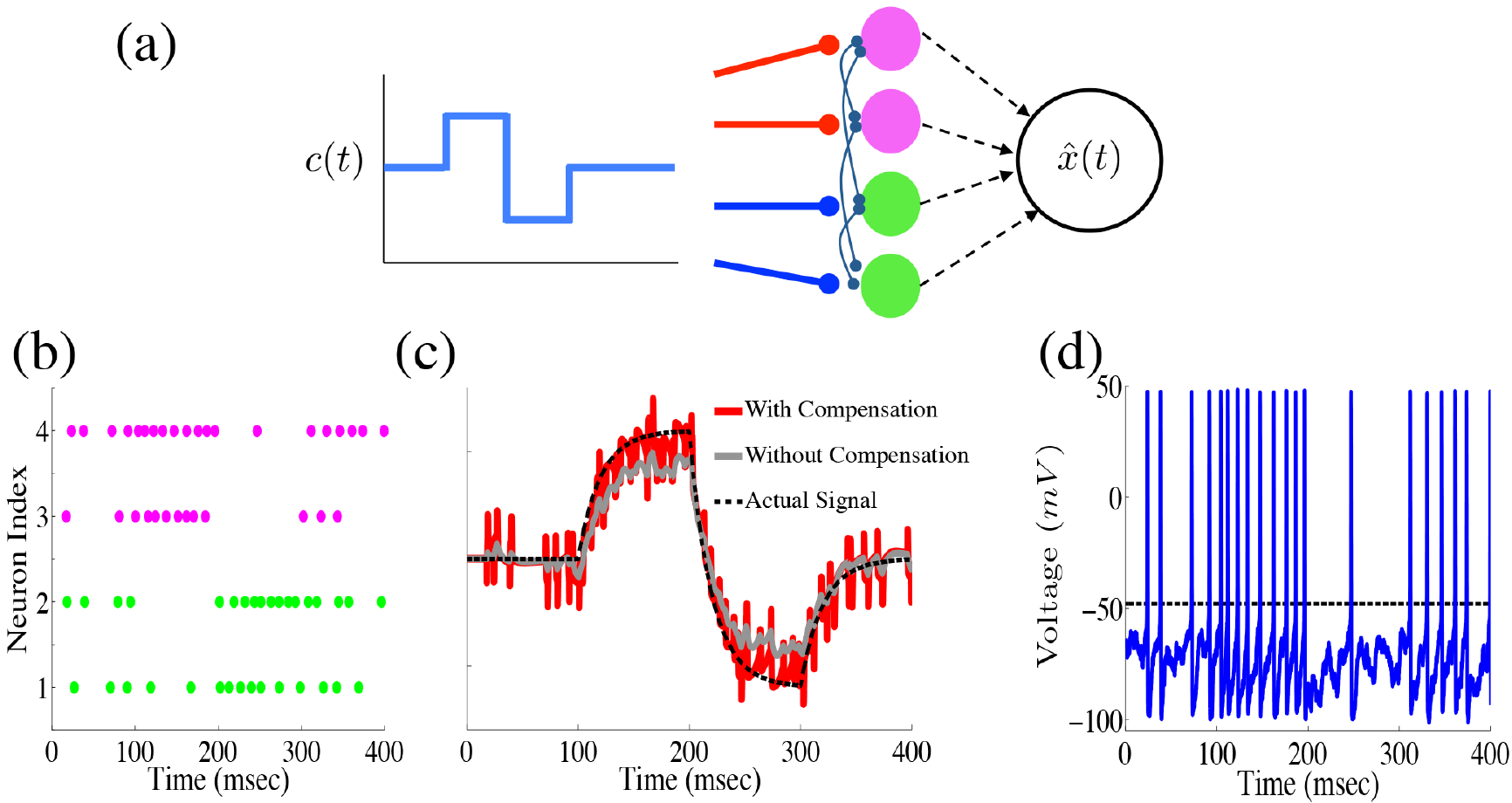}\\
{\bf Figure 3}
\end{center}


\clearpage
\newpage

\begin{center}
\includegraphics[scale=1]{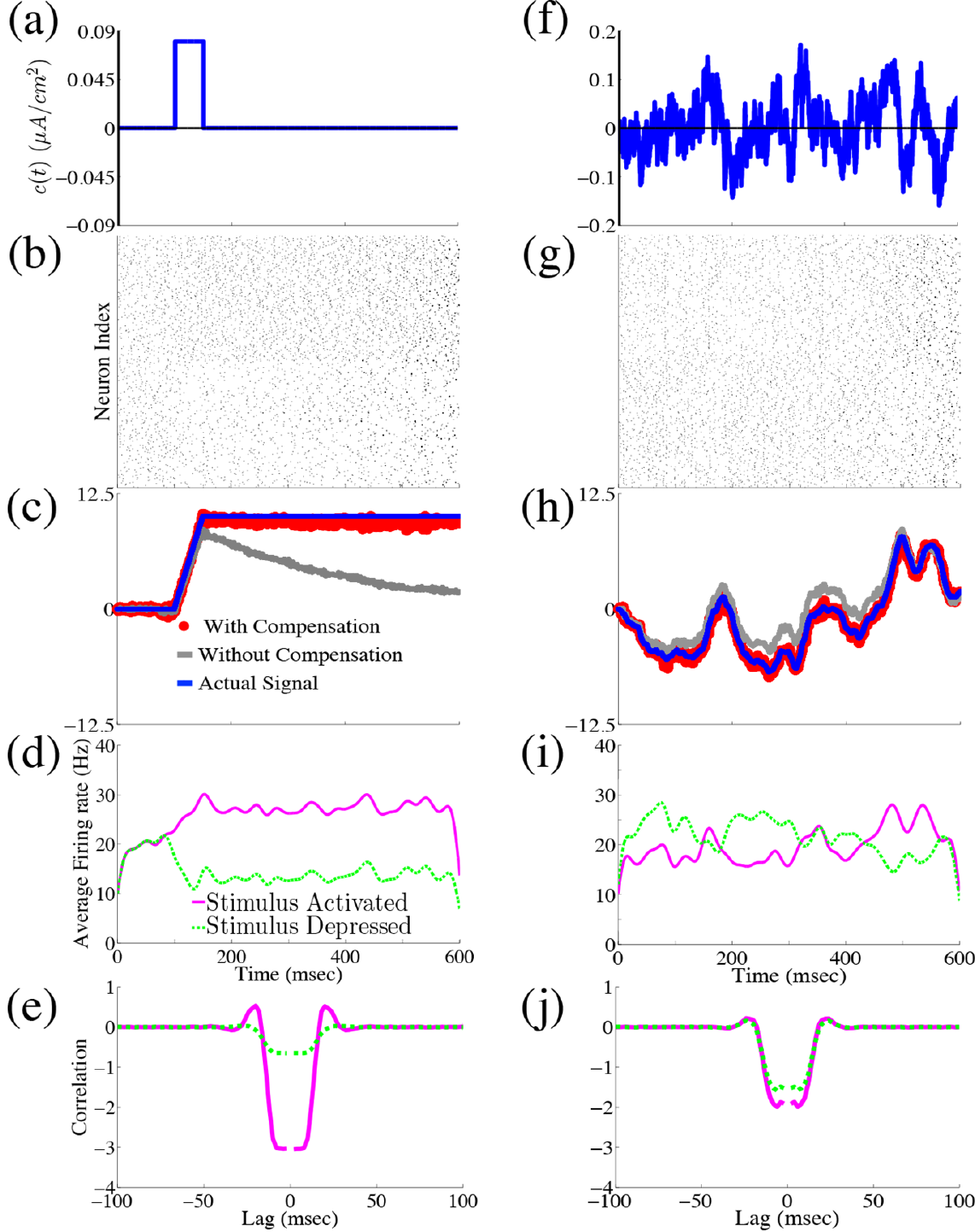}\\
{\bf Figure 4}
\end{center}


\clearpage
\newpage

\begin{center}
\includegraphics[scale=1]{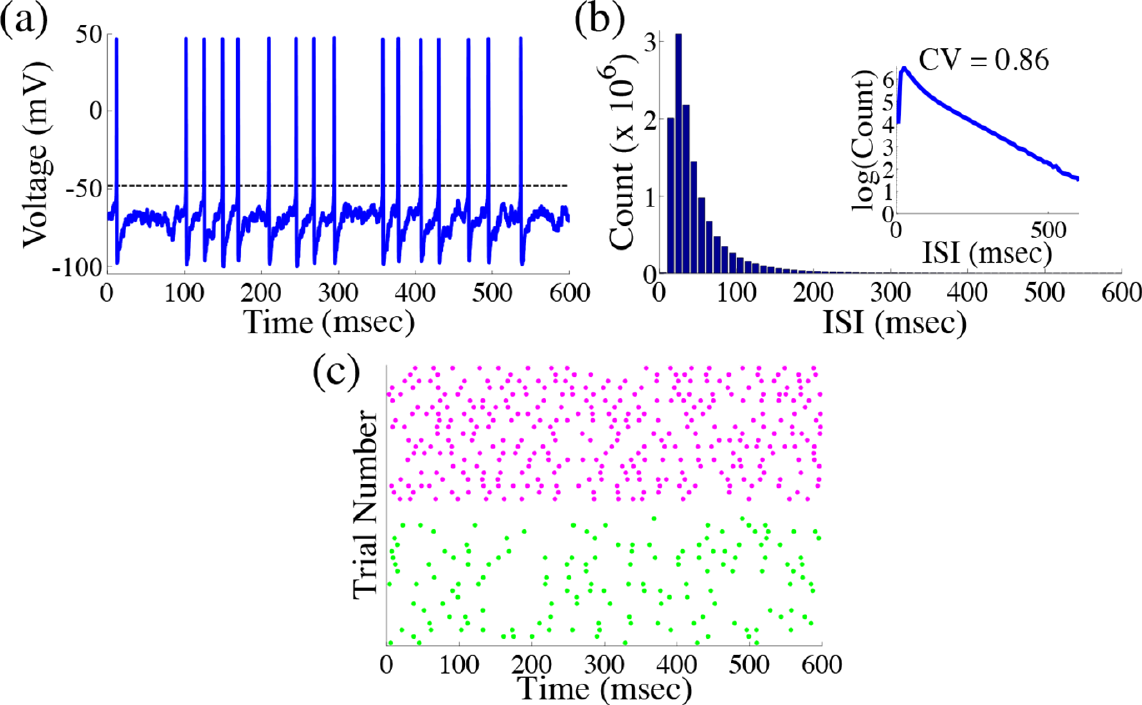}\\
{\bf Figure 5}
\end{center}


\clearpage
\newpage

\begin{center}
\includegraphics[scale=1]{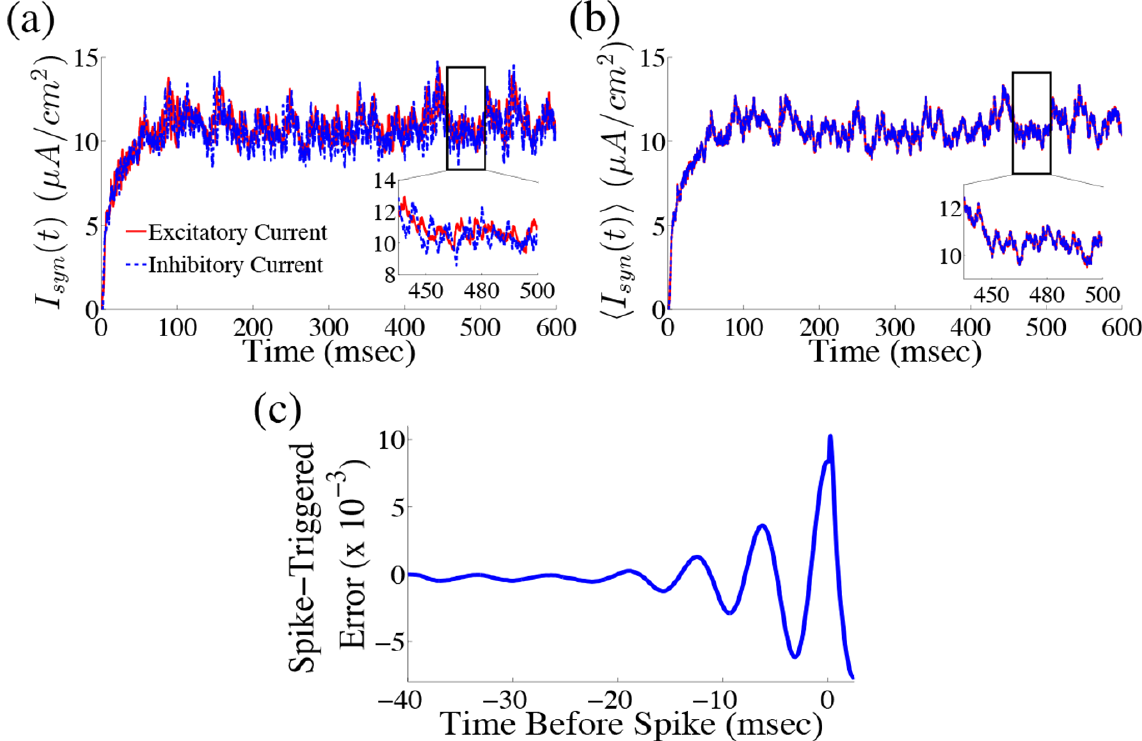}\\
{\bf Figure 6}
\end{center}


\clearpage
\newpage

\begin{center}
\includegraphics[scale=1]{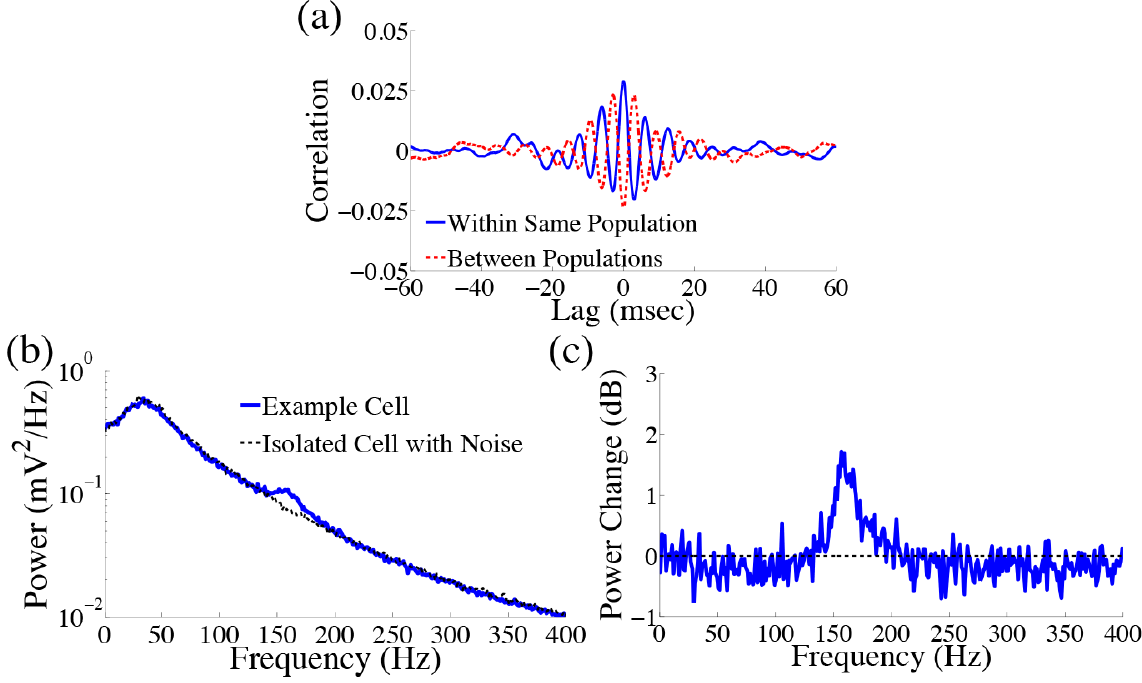}\\
{\bf Figure 7}
\end{center}


\clearpage
\newpage

\begin{center}
\includegraphics[scale=1]{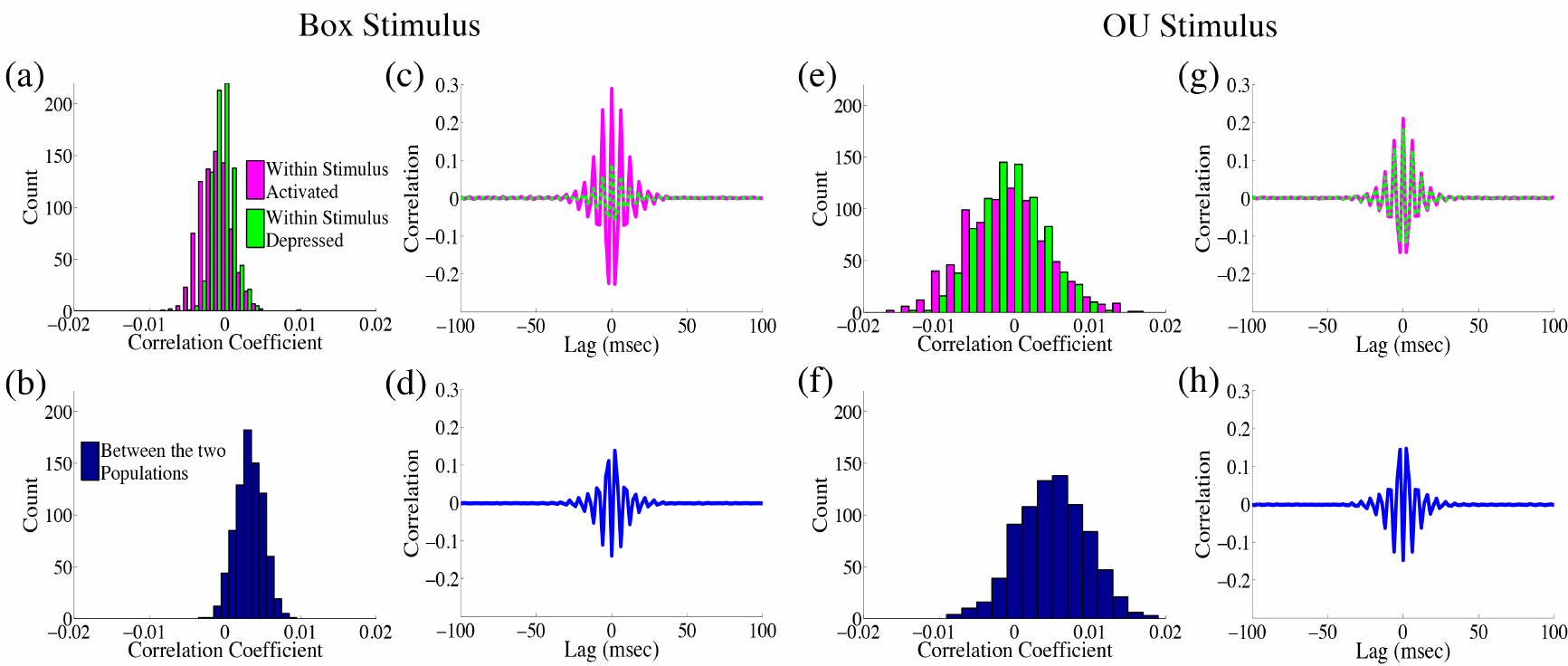}\\
{\bf Figure 8}
\end{center}


\clearpage
\newpage

\begin{center}
\includegraphics[scale=1]{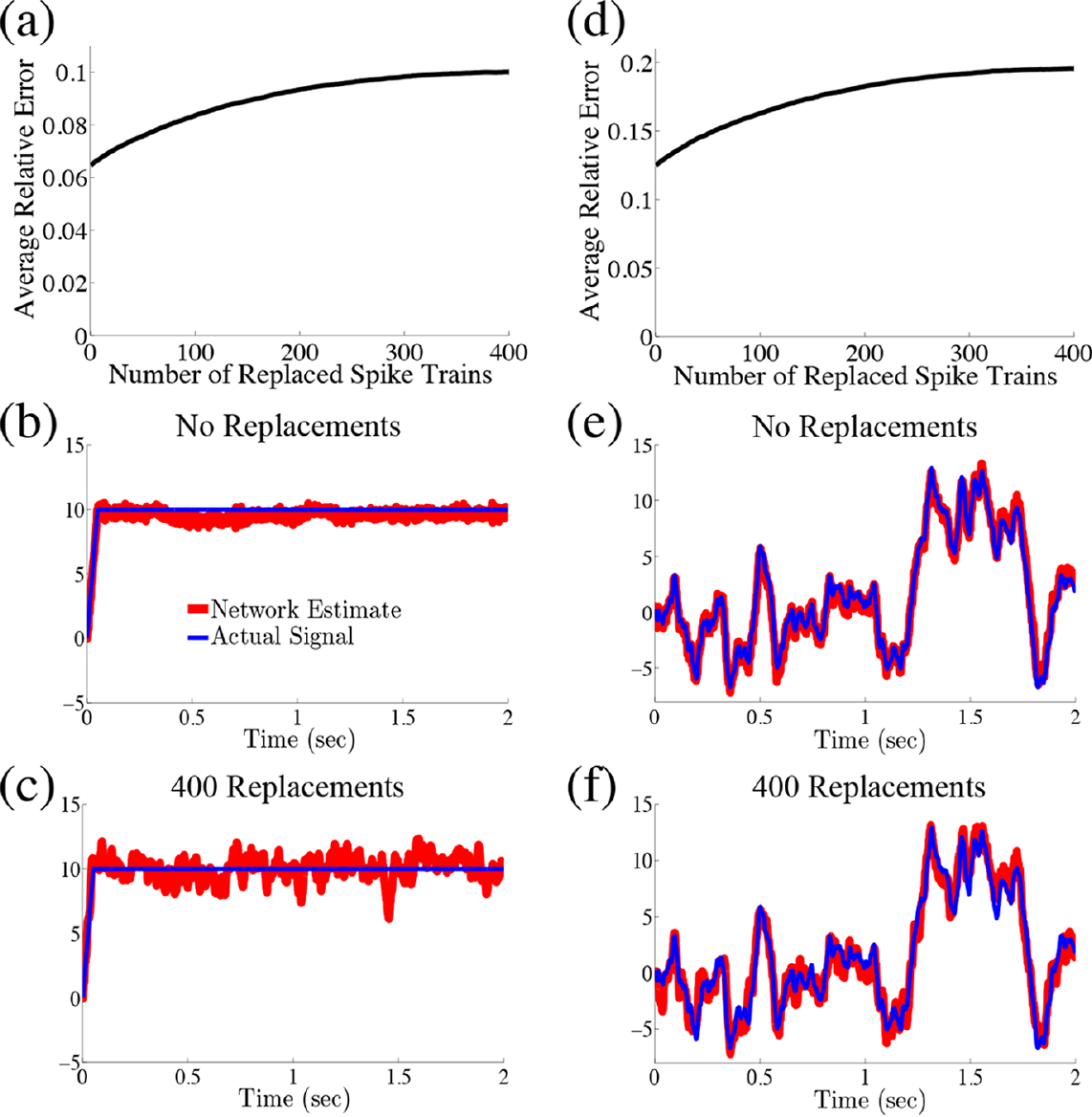}\\
{\bf Figure 9}
\end{center}


\clearpage
\newpage

\begin{center}
\includegraphics[scale=1]{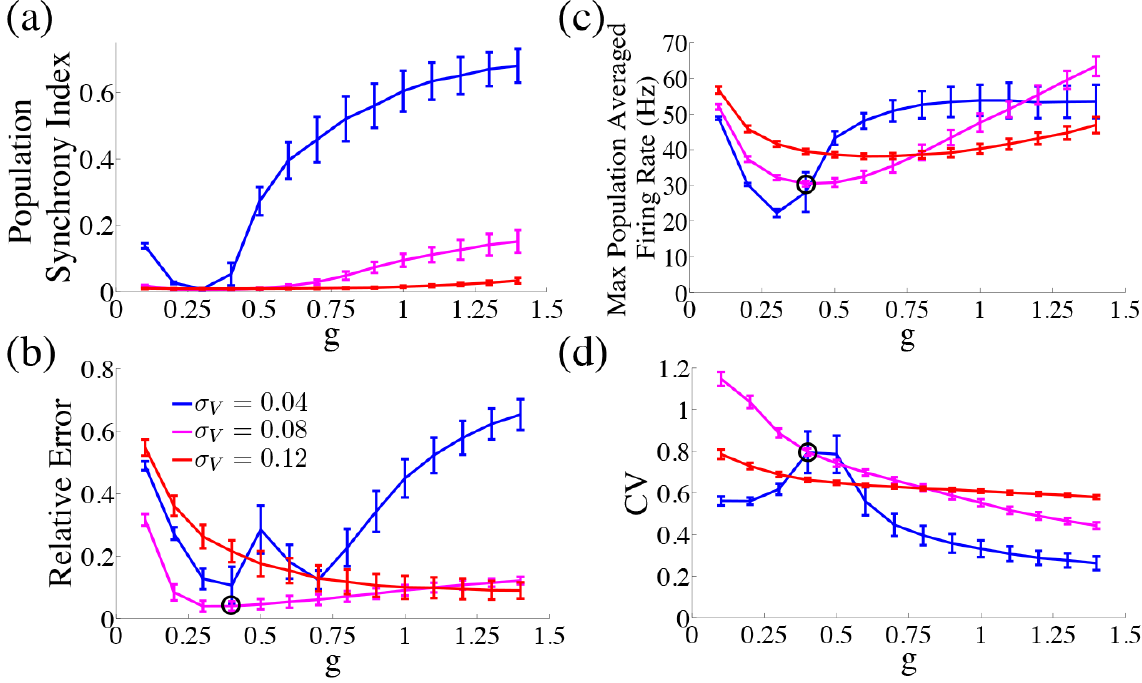}\\
{\bf Figure 10}
\end{center}


\clearpage
\newpage

\begin{center}
\includegraphics[scale=1]{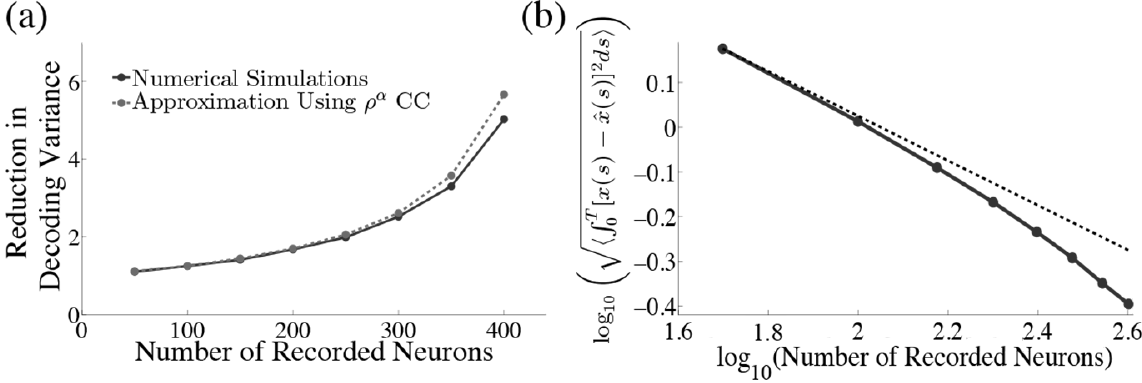}\\
{\bf Figure 11}
\end{center}


\clearpage
\newpage

\begin{center}
\includegraphics[scale=1]{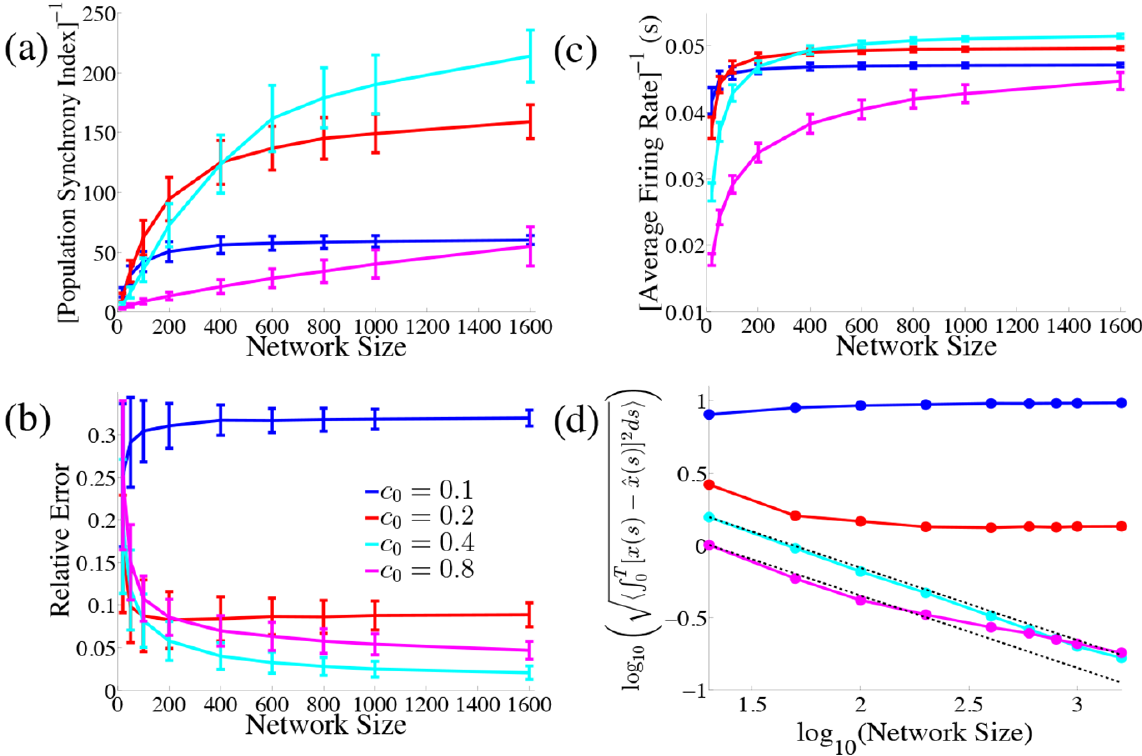}\\
{\bf Figure 12}
\end{center}


\clearpage
\newpage

\begin{center}
\includegraphics[scale=1]{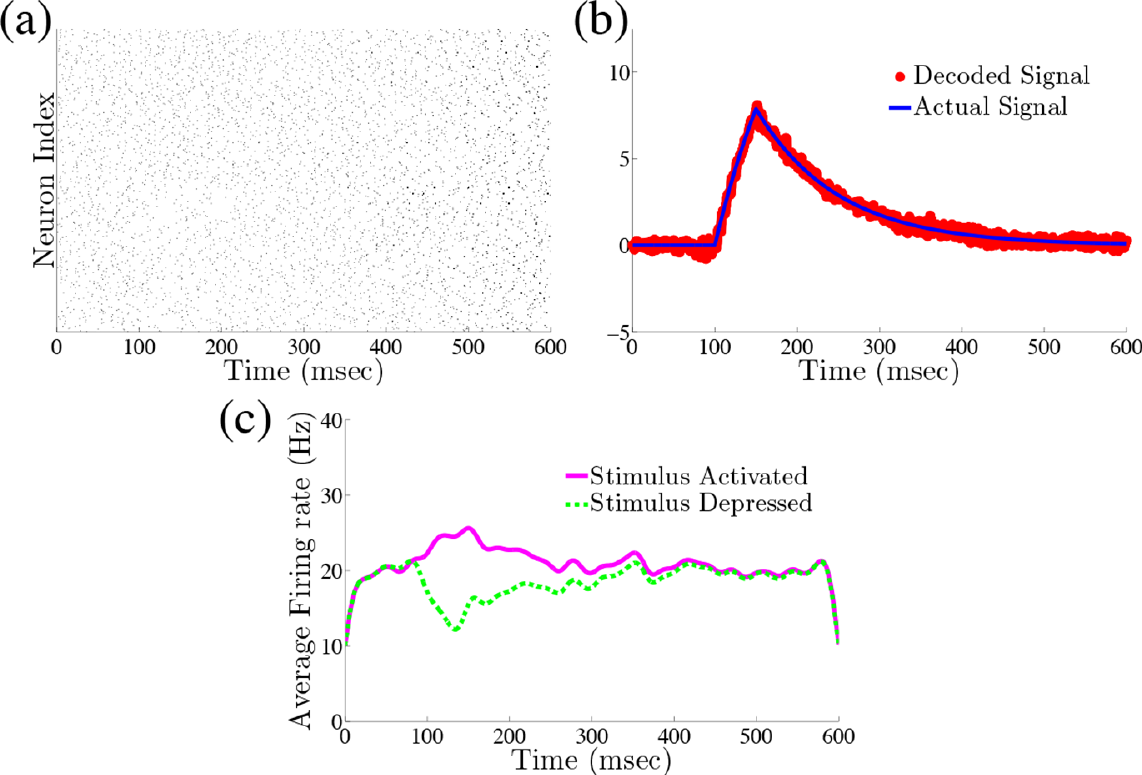}\\
{\bf Figure 13}
\end{center}


\clearpage
\newpage

\begin{center}
\includegraphics[scale=1]{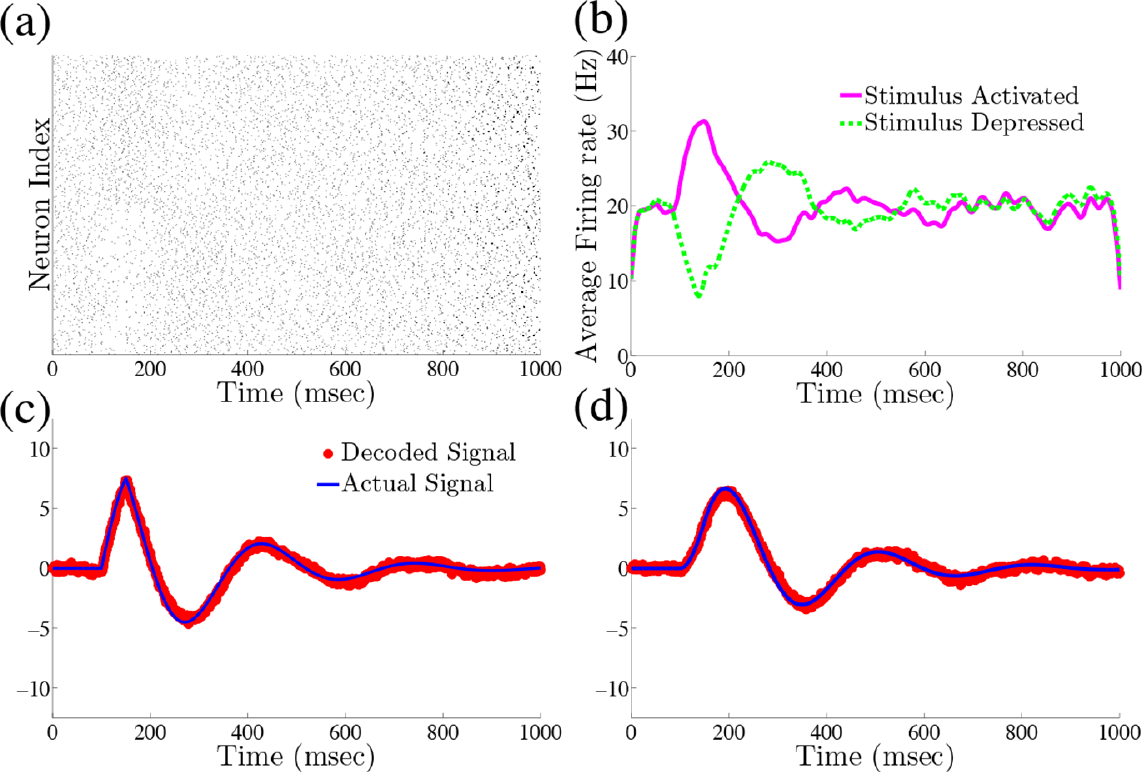}\\
{\bf Figure 14}
\end{center}


\end{document}